\documentclass[reprint,eqsecnum,floats,aps,amsmath,amssymb%
,footinbib,prd,showpacs]{revtex4-1}
\usepackage{amsmath,amsthm,latexsym,amssymb,amsfonts}
\usepackage{enumerate}
\usepackage{graphicx}
\usepackage{subfigure}
\usepackage{calc}

\usepackage[active]{srcltx}

\begin{document}

\title{Hybrid quantization of an inflationary universe}

\author{Mikel \surname{Fern\'andez-M\'endez}${}^{1}$}
\email{m.fernandez@iem.cfmac.csic.es}
\author{Guillermo A. \surname{Mena Marug\'an}${}^{1}$}
\email{mena@iem.cfmac.csic.es}
\author{Javier \surname{Olmedo}${}^{1}$}
\email{olmedo@iem.cfmac.csic.es}

\affiliation{
  ${}^{1}$Instituto de Estructura de la Materia, IEM-CSIC,
  Serrano 121, 28006 Madrid, Spain
}

\begin{abstract}
We quantize to completion an inflationary universe with small inhomogeneities
in the framework of loop quantum cosmology.
The homogeneous setting consists of a massive scalar field propagating in a
closed, homogeneous scenario. We provide a complete quantum description of the
system employing loop quantization techniques. After introducing small
inhomogeneities as scalar perturbations, we identify the true physical degrees
of freedom by means of a partial gauge fixing, removing all the local degrees of
freedom except the matter perturbations. We finally combine a Fock description
for the inhomogeneities with the polymeric quantization of the
homogeneous background, providing the quantum Hamiltonian constraint of the composed
system. Its solutions are then completely characterized, owing to the
suitable choice of quantum constraint, and the physical Hilbert
space is constructed. Finally, we consider the analog description for an
alternate gauge and, moreover, in terms of gauge-invariant quantities.
In the deparametrized model, all these descriptions are unitarily
equivalent at the quantum level.
\end{abstract}


\maketitle

\section{Introduction}

Several approaches attempt to combine the current classical description of the
gravitational interaction with the principles of the quantum theory. One of the
most promising candidates is the so-called Loop Quantum Gravity
(LQG)~\cite{lqg}, a non-perturbative, background-independent, canonical
quantization of General Relativity. Its application to cosmological models is
known as Loop Quantum  Cosmology (LQC)~\cite{lqc}. All the systems which have
been quantized in this framework show the same remarkable property: the
classical big-bang singularity is replaced by a quantum bounce. This is indeed
the case of massless scalar fields propagating in Friedmann-Robertson-Walker
(FRW) scenarios with flat~\cite{flat_lqc}, closed~\cite{apsv}, or open
topologies~\cite{v_open}, with cosmological constant~\cite{lambda_bp}, or even
in spacetimes with anisotropies~\cite{bianchi}.
Also inhomogeneous systems have been studied,
both perturbatively, considering tensor~\cite{tensor}, vector~\cite{vector}, and
scalar~\cite{scalar,scalar1} perturbations, and non-perturbatively,  as in the
case of Gowdy cosmologies in vacuo~\cite{gowdy,gowdy1,gowdy2} and with matter
content~\cite{gowdy_matt}. Considerable progress has been reached in situations in
which the inhomogeneities propagate as a field in an {\emph {effective}}
background, where quantum corrections in the geometry have been partly incorporated. The
analytic and numerical analysis carried out in Ref.~\cite{gowdy2} (treating the field classically)
showed that,
even in the presence of non-perturbative inhomogeneities (and disregarding the possibility of
extreme fine-tuning conditions), the quantum bounce persists. However, a more fundamental
description can be achieved when the field and the background are both treated quantum
mechanically, so that, in particular, their interaction is not limited to conform to any
particular effective dynamics~\cite{akl}. This issue has been already addressed
in Refs.~\cite{gowdy,gowdy1,gowdy_matt}, making use of the fact that Gowdy
scenarios with $T^3$ spatial topology, after a suitable gauge fixing, can be
interpreted as a scalar field propagating in a Bianchi~I cosmology. A completely
quantum description of those models was achieved through a hybrid
quantization scheme, proposed initially in Ref.~\cite{gowdy}, which combines LQC
quantization techniques for the homogeneous sector with a standard Fock
representation for the inhomogeneous degrees of freedom. In this approach, after a partial gauge fixing,
the spatial average of the Halmitonian constraint is imposed quantum mechanically,
and the physical Hilbert space is constructed out of its solutions.

Despite the variety of cosmological settings described above, more
attention needs to be drawn to inflationary scenarios, given their fundamental role in the
physics of the Early Universe~\cite{liddle}. Among all these scenarios, the case of a
massive scalar field propagating in an FRW spacetime is of special interest: it
is one of the simplest inflationary models keeping most of the required aspects
for the satisfactory understanding of our universe. Recently, it has been
possible to prove --at the LQC effective dynamics level-- that almost all of its
solutions provide enough inflation~\cite{as_slow}, solving the fine-tuning
problem arising in General Relativity. However, such  promising results do not
follow from any genuine quantum dynamics, owing to the lack of a complete
quantization of the system. On the other hand, inflationary universes provide a natural
framework for the development of primordial cosmological
perturbations~\cite{mukh}. The origin of those inhomogeneities could be explained
by the assumption that they stem from the early vacuum fluctuations of the
inflaton field. Assuming they were initially small, one can invoke perturbation
theory to treat them \cite{lif,bardeen}. As far as observations are involved, scalar
perturbations are the most interesting ones, since they are those that leave
stronger imprints, e.g., in the cosmic microwave background. Thus, they have
been observed with high precision by the Wilkinson Microwave Anisotropy Probe
(WMAP)~\cite{last_WMAP}. Though, even when this type of inhomogeneities has been
partially analyzed in LQC inflationary models~\cite{scalar,scalar1}, it is
essential to confront the results achieved so far with those that could be
obtained from a complete quantization of the perturbed system, without any
restriction to a particular effective background.

To approach this problem, in this article we will carry out a complete
quantization of an inflationary cosmological model with small inhomogeneities.
In a first step, we will consider a homogeneous massive scalar field propagating
in a closed FRW scenario~\cite{comment1}. A quantum Hamiltonian constraint is
provided and its solutions are characterized. Specifically, it is noted that
they are completely determined by their initial data on the minimum volume
section allowed by the discretization of the geometry. The physical
Hilbert space can be constructed out of this space of initial data just by
equipping it with a suitable inner product. Once the homogeneous setting is
established, we will introduce small inhomogeneities around the homogeneous
solutions by means of perturbation theory. We will carry out a gauge fixing
(adopting the longitudinal gauge), and a canonical transformation which includes
the scaling of the matter perturbation by the FRW scale factor. The aim of such
a transformation is to achieve a formulation of the system in which the
dynamical behavior of the matter perturbation approaches, in the ultraviolet limit, the
one of a scalar field propagating in a static spacetime with an effective mass
owing to the interaction with the FRW background. Next, we will proceed to
quantize the system employing a hybrid approach, i.e., we will apply a
polymeric quantization to the homogeneous sector while adopting a Fock
representation for the inhomogeneities. The problem that then arises concerns
the selection of a particular Fock quantization, as we have at our disposal
infinite inequivalent representations. Fortunately, the recent uniqueness
results for fields in non-stationary spacetimes~\cite{uniq} motivate us to pick
up the family of Fock representations in which the vacuum state is invariant
under the group of spatial isometries --namely, the $SO(4)$ group-- and where the
classical dynamics of the field (obtained after deparametrization)
is implemented by a unitary quantum operator. It
has been possible to prove~\cite{fmo} that this family is indeed a unique
unitary equivalence class, as in the situations treated in Ref.~\cite{uniq}.
This result is closely tied to the previously mentioned
canonical transformation, since a different (non-trivial) scaling of the field
prevents the unitary implementation of its corresponding dynamics.
Finally, we will formally provide the quantum Hamiltonian constraint of the system
and its solutions, and endow them with a Hilbert space structure, thus obtaining
the physical Hilbert space.

In order to show the robustness of the treatment and avoid problems related with the choice of
a specific gauge, we repeat the
procedure for an alternate gauge, in which the spatial metric is homogeneous.
Moreover, in both situations we relate the modes of the resulting canonical pair of
fields with a set of Bardeen potentials~\cite{bardeen}, which are gauge invariants of the model,
and prove that the canonical transformations relating our fundamental fields with these
potentials are unitarily implementable in the quantum theory.

This article is organized as follows. In Sec.~\ref{sec:hom_sys} we provide a
classical description and the complete quantization of the homogeneous system.
The inhomogeneities are included at the classical level in
Sec.~\ref{sec:inhomog_class}, and a complete hybrid quantization of the system
is carried out in Sec.~\ref{sec:hybrid_quant} for a particular gauge choice. In
Sec.~\ref{sec:gauge_B}, we repeat the same analysis for an alternate gauge
fixing. A description in terms of Bardeen potentials can be found in
Sec.~\ref{sec:bardeen}. A summary with the main conclusions is given in
Sec.~\ref{sec:conclusions}. Finally, to make the paper self-contained, four appendixes
have been included.

\section{The homogeneous and isotropic system}\label{sec:hom_sys}

In this section, we will deal with the homogeneous and isotropic setting, which consists of a
massive scalar field propagating in a closed FRW spacetime. We will give a
classical description of the corresponding constrained system and proceed to
quantize the model to completion, determining the physical Hilbert space.
Concerning the technical aspects of the quantization, we will mainly follow
Ref.~\cite{apsv}.

\subsection{Classical system}\label{sec:hom_class}

The FRW spacetime under consideration can be foliated in spatial sections
isomorphic to three-spheres. On them, we choose as fiducial metric the standard round metric of
unit radius (instead of radius $a_0 = 2$ as in
Ref.~\cite{apsv}), with fiducial volume $l_0^3=2\pi^2$.

The degrees of freedom of the matter sector --without taking into account
constraints-- are a massive scalar field $\phi$ and its canonically conjugate
momentum $p_\phi$, while the geometry will be described in terms
of an Ashtekar-Barbero connection $A^i_a=c \,\,{}^{0}\omega_a^i/l_0$ and a
densitized triad $E^a_i=p\sqrt{\Omega}\,\,{}^{0}e_i^a/l_0^{2}$, with $i$ being
the internal $SU(2)$ index and $a$ the spatial one, both running from 1 to 3.
Here, $\Omega_{ab}={}^{0}\omega_a^i\,\,\,{}^{0}\omega_b^i$ is the fiducial
metric of the spatial sections, and $\Omega$ its determinant. The classical algebra is given by
$\{\phi,p_\phi\}=1$ and $\{c,p\}=8\pi G\gamma/3$, where $G$ and $\gamma$ are the
Newton constant and the Immirzi parameter, respectively.

Owing to the homogeneity and isotropy, only the Hamiltonian constraint
$\mathbb{H}=\bar{N}_0C_0/(16\pi G )$ remains, where $\bar{N}_0$ is the (homogeneous) lapse function,
\begin{equation}
\label{eq:homclassconst}
C_0 =-\frac{6\sqrt{|p|}}{\gamma^2}[(c-l_0)^2+\gamma^2l_0^2] + \frac{8\pi G}{V}\left(p_\phi^2+m^2V^2\phi^2\right),
\end{equation}
and $V=|p|^{3/2}$.

\subsection{Quantum system}\label{sec:hom_quant}

In the quantization of the system, we choose a standard Schr\"odinger
representation for the scalar field, whose kinematical Hilbert space is
$\mathcal{H}^{\rm mat}_{\rm kin}=L^2(\mathbb{R},d\phi)$. For the geometry, we
apply a polymeric quantization \cite{lqc}. The basic variables in LQC are fluxes of
densitized triads through square surfaces, enclosed by four geodesic edges,
which are basically given by $p$, and holonomies of the connection along integral
curves of the fiducial triads of fiducial length $\mu l_0$. The gravitational
part of the kinematical Hilbert space is $\mathcal{H}^{\rm grav}_{\rm
kin}=L^2(\mathbb{R}_{\rm Bohr},d\mu_{\rm Bohr})$, where $\mathbb{R}_{\rm Bohr}$
is the Bohr compactification of the real line, and $d\mu_{\rm Bohr}$ is the
natural Haar measure associated with it. Then, the kinematical Hilbert space of the whole system is
the tensor product $\mathcal{H}^{\rm mat}_{\rm kin}\otimes  \mathcal{H}^{\rm
grav}_{\rm kin}$.

To promote the Hamiltonian constraint to a quantum operator, its gravitational
part must be written in terms of holonomies of the connection $A_a^i$. We will
essentially follow  the procedure of Ref.~\cite{apsv}, where the curvature
tensor constructed out of $su(2)$-connections is regularized. As a first step,
one constructs holonomies along the edges of a closed square, properly selected
by considering four integral curves along alternating left and right invariant
vector fields (left and right invariant vector fields commute), well adapted to
the fiducial structures. Then, the curvature operator is basically replaced by a
circuit of holonomies around a square enclosing a non-vanishing area $\Delta$,
determined by the infrared spectrum of the area operator defined in LQG
(classically, the local curvature would be recovered in the limit $\Delta\to
0$). As in Ref.~\cite{apsv}, we adhere to the {\it improved dynamics}
scheme, in which the fiducial length of the edges of the considered square
is given by the function $\bar{\mu}=\sqrt{\Delta/p}$, up to a factor $l_0$. In the
triad representation, one can find a basis of normalizable states $|v\rangle$ in
$\mathcal{H}^{\rm grav}_{\rm kin}$ (with $v\in \mathbb{R}$), on which the action
of the matrix elements of the holonomies is $\hat
N_{\bar{\mu}}|v\rangle=|v+1\rangle$, while $\hat p|v\rangle={\rm sgn}(v)(2\pi
\gamma G \hbar\sqrt{\Delta} |v|)^{2/3}|v\rangle$, with $\hbar$ being the Planck
constant.

Finally, we adopt the following operator as quantum
Hamiltonian constraint:
\begin{eqnarray}\label{eq:hom_quantum_const}
\hat{C}_0&=&\widehat{\left[\frac{1}{V}\right]}^{1/2}\bigg[8\pi G \big(\hat{p}_\phi^2+m^2\hat{V}^2\hat{\phi}^2\big)-\frac{6}{\gamma^2}\widehat{\Omega}^2\\ \nonumber
&-&\frac{6}{\gamma^2}\bigg\{(1+\gamma^2)l_0^2
\hat{V}^{4/3}-\frac{\hat{V}^2}{\Delta}\sin^2(\hat{\bar{\mu}}l_0)\bigg\}\bigg]\widehat{\left[\frac{1}{V}\right]}^{1/2}\!\!\!\!\!,
\end{eqnarray}
where $\hat{V}$ represents the volume $V=|p|^{3/2}$, and we have introduced the inverse of this operator, defined as
\begin{eqnarray}\nonumber
\widehat{\left[\frac{1}{V}\right]}&=&\widehat{{\rm sgn}(v)}\hat V\bigg[\frac{3}{4\pi \gamma G \hbar
\sqrt{\Delta}}(\hat{N}_{-\bar{\mu}}\hat V^{1/3}\hat{N}_{\bar{\mu}}\\
 &-&\hat{N}_{\bar{\mu}}\hat V^{1/3}\hat{N}_{-\bar{\mu}})\bigg]^{3}.
\end{eqnarray}
In addition,
\begin{eqnarray}
\label{eq:quantumomega}  &\widehat{\Omega} &=
\frac{1}{4i\sqrt{\Delta}} \hat{V}^{1/2}\\ \nonumber  &\times&
\Big[\widehat{{\rm sgn}(v)}
\left(e^{-i\frac{\hat{\bar{\mu}}l_0}{2}}\hat{N}_{2\bar{\mu}}e^{-i\frac{\hat{\bar{\mu}}l_0}{2}}-
e^{i\frac{\hat{\bar{\mu}}l_0}{2}}\hat{N}_{-2\bar{\mu}}e^{i\frac{\hat{\bar{\mu}}l_0}{2}}\right)
\\ \nonumber &+&\left(e^{-i\frac{\hat{\bar{\mu}}l_0}{2}}\hat{N}_{2\bar{\mu}}e^{-i\frac{\hat{\bar{\mu}}l_0}{2}}-
e^{i\frac{\hat{\bar{\mu}}l_0}{2}}\hat{N}_{-2\bar{\mu}}e^{i\frac{\hat{\bar{\mu}}l_0}{2}}\right)\widehat{{\rm sgn}(v)}\Big]
\hat{V}^{1/2}.
\end{eqnarray}

The constraint $\hat{C}_0$ annihilates the state $|v=0\rangle$ and leaves
invariant its orthogonal complement. Therefore, we can decouple the state
corresponding to the classical singularity, i.e., $|v=0\rangle$, and restrict
the study to its orthogonal complement, that will be denoted from now on by
$\widetilde{\mathcal{H}}^{\rm grav}_{\rm kin}$. Besides, with the usual definition
of $\sin(\hat{\bar{\mu}}l_0)$ (see Ref.~\cite{apsv}), $\widehat{\Omega}^2$ is
the only operator in the constraint with a non-diagonal action on the $v$-basis of
$\widetilde{\mathcal{H}}^{\rm grav}_{\rm kin}$. It only relates states with
support on isolated points separated by a constant step of $4$ units in the
label $v$ and, moreover, different orientations of the triad are decoupled. In
conclusion, only states with support on semilattices
$\mathcal{L}_{\varepsilon}^\pm=\{v=\pm(\varepsilon+4n);\,n\in \mathbb{N}\}$ are
related by the action of $\widehat{\Omega}^2$, where $\varepsilon\in(0,4]$ is a
continuous parameter proportional to the minimum value of the physical volume of
the system in the sector under consideration. Let us emphasize that those
sectors are preserved by all the  operators of physical interest considered in
this article. Consequently, they can be interpreted as superselection sectors,
denoted from now on as  $\mathcal{H}^{\varepsilon}_\pm$.

In the following, and without loss of generality, we will restrict our study to
$\mathcal{H}^{\varepsilon}_+$. Besides, we will apply a unitary
transformation $\hat U=e^{i l_0\hat{h}( v)}$ on  $\mathcal{H}^{\varepsilon}_+$,
where $h(v)$ is defined in Appendix~\ref{appendixA}, in complete parallelism
with the analyses carried out in Ref.
\cite{apsv}. This unitary transformation maps $\widehat{\Omega}^2$ into
\begin{eqnarray}\label{eq:flat_omega}
\widehat\Omega_0^2=\hat U \widehat\Omega^2 \hat U^{-1},
\end{eqnarray}
which is equivalent to the analog operator for a scenario with flat topology
\begin{eqnarray}
\label{eq:quantumomega_flat}  &\widehat{\Omega}_0 &=
\frac{1}{4i\sqrt{\Delta}}\hat{V}^{1/2}\Big[\widehat{{\rm sgn}(v)}\big(
\hat{N}_{2\bar{\mu}}-\hat{N}_{-2\bar{\mu}}\big)\\ \nonumber
&+&\big(\hat{N}_{2\bar{\mu}}-\hat{N}_{-2\bar{\mu}} \big)\widehat{{\rm
sgn}(v)}\Big]\hat{V}^{1/2},
\end{eqnarray}
just like the operator suggested in Ref.~\cite{mmo}, but without
inverse volume corrections (see Ref.~\cite{mop} for additional
information).

Hence, the gravitational part of the quantum Hamiltonian constraint is
finally encoded by
\begin{eqnarray}
\hat{C}_{\rm grav}&=&
\widehat{\left[\frac{1}{V}\right]}^{1/2} \hat{\mathcal{C}}_{\rm grav} \widehat{\left[\frac{1}{V}\right]}^{1/2},
\end{eqnarray}
with
\begin{equation}
\hat{\mathcal{C}}_{\rm grav}=-
\frac{6}{\gamma^2} \bigg\{\widehat{\Omega}_0^2-\frac{\hat{V}^2}{\Delta}\sin^2(\hat{\bar{\mu}}l_0)+
(1+ \gamma^2)l_0^2
\hat{V}^{4/3}\bigg\}.
\end{equation}

As an aside, let us comment that, following the arguments of Ref.~\cite{apsv} (applied there to a related operator, $\hat \Theta$), and noticing that the superselection sectors are now semi-lattices (instead of full lattices as in that work), it seems reasonable to
admit that the restriction of $\hat{\mathcal{C}}_{\rm grav}$ to
$\mathcal{H}^{\varepsilon}_+$ provides a self-adjoint operator with a non-degenerate
discrete spectrum.

The solutions $(\Psi|$ annihilated by the constraint $\hat{C}_0$ can be defined
in the dual of a dense set in $ \mathcal{H}^{\rm mat}_{\rm kin}\otimes
\mathcal{H}^{\varepsilon}_+$, e.g., that of the product of the span of the $v$-basis and the functions of $\phi$ with compact support.
It is straightforward to see that
the coefficients $\Psi(v,\phi)=(\Psi|v,\phi\rangle$ satisfy a difference
equation that resembles an evolution equation with $v$ playing the role of an
internal time. Besides, one can check that the initial data
$\Psi(\varepsilon,\phi)$ determines the whole solution at any volume
$v=\varepsilon + 4n, \,\,\forall n\in \mathbb{N}$, in the considered semilattice. Therefore, we can
identify these initial data with the solutions, and
complete the space of such data with an adequate inner product in order to reach the physical Hilbert
space \cite{note}. In this way, the physical Hilbert space can be taken as
$\mathcal{H}^{\rm phys} = L^2(\mathbb{R},d\phi)$.

\section{Inhomogeneities: classical description}\label{sec:inhomog_class}

In this section we will introduce small inhomogeneities in the model, treating them as perturbations.
Specifically, we will expand the Hamiltonian constraint up to second
order in those perturbations. We will follow the same classical approach as in Ref.~\cite{hh}.
Besides, we will carry out a gauge fixing in order to remove the unphysical degrees of freedom.
Finally, we will introduce a suitable canonical transformation to prepare the system for its hybrid
quantization, performed in Sec.~\ref{sec:hybrid_quant}.

\subsection{Perturbations around classical homogeneous solutions.}\label{sec:inhom_class_pert}

We now briefly revisit the inclusion of inhomogeneities in our cosmological model as perturbations
of the homogeneous system. For convenience, we introduce a perturbative parameter $\epsilon$,
making proportional to it each of the inhomogeneous
corrections to the geometry and the matter field. Using this parameter, we carry out a
perturbative expansion in the action, expressed in Hamiltonian form, and truncate the expansion at
second order. Furthermore, for simplicity, we will only consider scalar
perturbations, something which is consistent at the studied perturbative order because they are decoupled from
genuine vector and tensor perturbations of the system in this approximation \cite{hh}.
Our focus on scalar perturbations is mainly motivated by their
relevant role in observational cosmology. In addition,
we adopt a natural expansion of the inhomogeneities in terms of modes of the Laplace-Beltrami operator in the three-sphere,
namely, the (hyper-)spherical harmonics in
$S^3$:  $Q^n$, $P_{a}^n$, and $P_{ab}^n$ (see Appendix~\ref{appendixB}), where we recall that Latin indices from the beginning of the alphabet denote spatial indices, and $n\in \mathbb{N}^+$ is here \cite{comment1b} a positive integer that labels the eigenvalues of the Laplace-Beltrami operator [see Eq. \eqref{Qn}]. Let us then write
\begin{eqnarray}\label{eq:lapse}
N&=&\sigma N_0\Big(1+\sqrt{2}\pi\epsilon\sum_n g_nQ^n\Big),\\
\label{eq:shift}
N_a&=& \sigma^2 e^{\alpha} \sqrt{2}\pi\epsilon\sum_n k_nP_a^n,\\
\label{eq:spatial_metric}
h_{ab}&=& \sigma^2 e^{2\alpha} \Big[\Omega_{ab}+2^{3/2}\pi\epsilon\sum_n (a_n Q^n \Omega_{ab}\nonumber
\\ &+&3b_nP^n_{ab})\Big],\\
\label{eq:matter_field}
\Phi&=&\frac{1}{\sigma}\Big[\frac{\varphi}{\sqrt{2}\pi}+\epsilon\sum_nf_n Q^n\Big].
\end{eqnarray}
In these formulas, $\sigma^2=4\pi G/(3l_0^3)$, and the unperturbed scalar field has mass $m=\tilde{m}/\sigma$.
In order to facilitate comparison with the analysis of Ref. \cite{hh}, as well as to apply the results of Ref. \cite{fmo}, we employ here the variables $\alpha$, $\varphi$, and their corresponding momenta $\pi_{\alpha}$ and $\pi_\varphi$, to describe the homogeneous sector. In the absence of inhomogeneities (i.e., in the unperturbed system), the relation with the homogeneous
variables used in Sec.~\ref{sec:hom_class} is given by
\begin{eqnarray}\label{changehomo}
|p|&=&l_0^{2}\sigma^{2}e^{2\alpha},\quad\quad  p (c -l_0)= - \gamma l_0^{3} \sigma^{2} \pi_\alpha, \\ \label{changehomo2}
\phi&=&\frac{\varphi}{l_0^{3/2}\sigma},\quad \quad p_\phi=l_0^{3/2}\sigma \pi_\varphi.
\end{eqnarray}
Note that the correspondence between $\alpha\in \mathbb{R}$ and the flux variable $p$ is one-to-one, e.g., in the union of all the
superselection sectors $\mathcal{H}^{\varepsilon}_+$, to which we are particularizing our discussion.

For the inhomogeneities, on the other hand, the canonically
conjugate momenta of the coefficients $a_n$, $b_n$, and $f_n$ will be denoted by
$\pi_{a_n}$, $\pi_{b_n}$, and $\pi_{f_n}$, respectively, while $g_n$ and $k_n$
play the role of Lagrange multipliers associated with two linear constraints.

Truncating the action at quadratic order in $\epsilon$, we obtain
the total Hamiltonian:
\begin{equation}
\mathbb{H}=N_0\bigg[H_0+\epsilon^2\sum_n\big(H_2^n+g_nH_{|1}^n\big)\bigg]+\epsilon^2\sum_nk_n H_{\_\!\_1}^n,
\end{equation}
where $H_0$ corresponds to the Hamiltonian constraint of the homogeneous sector,
\begin{equation}\label{eq:hom_C0}
H_0=\frac{e^{-3\alpha}}{2}(-\pi_\alpha^2+\pi_\varphi^2+e^{6\alpha}\tilde m^2\varphi^2-e^{4\alpha}),
\end{equation}
and for each mode $n$ we have
\begin{widetext}
\begin{eqnarray}\label{eq:scala_const2}
H_2^n&=& \frac{e^{-3\alpha}}{2}\Bigg[\left(\frac12a_n^2+10\frac{n^2-4}{n^2-1}b_n^2\right)\pi_\alpha^2+ \left(\frac{15}2a_n^2+6\frac{n^2-4}{n^2-1}b_n^2\right)\pi_\varphi^2- \pi_{a_n}^2+  \frac{n^2-1}{n^2-4}\pi_{b_n}^2 +\pi_{f_n}^2 -6a_n\pi_{f_n}\pi_\varphi\nonumber\\
&+&\big(2a_n\pi_{a_n}+8b_n\pi_{b_n}\big)\pi_\alpha- e^{4\alpha}\bigg\{\frac13\big(n^2-\tfrac52\big)a_n^2+\frac13
\big(n^2-7\big)\frac{n^2-4}{n^2-1}b_n^2+\frac23\big(n^2-4\big)a_nb_n-\big(n^2-1\big)f_ n^2\bigg\} \nonumber\\
&+&e^{6\alpha}\tilde m^2\left(\frac32\varphi^2a_n^2-6\frac{n^2-4}{n^2-1}\varphi^2b_n^2+f_n^2+6\varphi a_nf_n\right)\Bigg],\\\label{eq:scala_const1}
H_{|1}^n &=& \frac{e^{-3\alpha}}{2}\Bigg[2(\pi_\varphi\pi_{f_n}-\pi_\alpha\pi_{a_n})-(\pi_\alpha^2+3 \pi_\varphi^2)a_n-\frac23e^{4\alpha}\Big\{\big(n^2+\tfrac12\big)a_n+\big(n^2-4\big)b_n\Big\}+e^{6\alpha}\tilde m^2\varphi\big(2f_n+3\varphi a_n\big)\Bigg],\,\,\,\,\,\,\,\,\,\,\,\,\\\label{eq:diffeo_const}
H_{\_\!\_1}^n &=& \frac{e^{-\alpha}}{3}\Bigg[\left(a_n+4\frac{n^2-4}{n^2-1}b_n\right)\pi_\alpha-\pi_{a_n}+\pi_{b_n}+3f_n \pi_\varphi\Bigg].
\end{eqnarray}
\end{widetext}
Here, $H_2^n$ and $H_{|1}^n$ are the scalar constraints quadratic and linear in the
perturbations, respectively, and $H_{\_\!\_1}^n$ is the diffeomorphism constraint, also
linear in the inhomogeneities. Finally, note that, substituting $\bar{N}_0=\sigma N_0$, the homogeneous constraint \eqref{eq:homclassconst}
reproduces $16 \pi G H_0$ as given in Eq. \eqref{eq:hom_C0},
under the changes \eqref{changehomo} and \eqref{changehomo2}.

\subsection{Gauge fixing}\label{sec:gauge_fixA}

For each mode $n=3,4,5,...$, there are three dynamical degrees of freedom in the
inhomogeneous sector: $a_n$, $b_n$, and $f_n$, and two linear constraints:
$H_{|1}^n$ and $H_{\_\!\_1}^n$. The modes
$n=1$ and $n=2$, on the other hand, will be treated independently, since they are in fact pure gauge.
Hence, for $n \geq 3$, two non-physical degrees of freedom can be removed by
means of a gauge fixing. Let us consider the longitudinal gauge, which is fixed
by imposing the conditions
\begin{eqnarray}\label{eq:gauge_A}
b_n= 0,\quad \Pi_{n}= \pi_{a_n}-\pi_\alpha a_n-3\pi_\varphi f_n=0.
\end{eqnarray}
These conditions provide an admissible gauge since they are second class
with the linear momentum and Hamiltonian constraints. Besides, we
can find values of $k_n$ and $g_n$ for which these gauge-fixing conditions are stable in the evolution. A simple computation yields that, on the gauge-fixed section,
\begin{eqnarray}\label{eq:gauge_A_bn}
\{b_n,\mathbb{H}\}&=&0 \Leftrightarrow k_n= -3\frac{n^2-1}{n^2-4}N_0e^{-2\alpha}\pi_{b_n},\\ \label{eq:gauge_A_Pin}
\{\Pi_n,\mathbb{H}\}&=&0 \Leftrightarrow g_n= -a_n.
\end{eqnarray}
Finally, if the conditions $H_{|1}^n =H_{\_\!\_1}^n= 0 $ are satisfied, we can
eliminate $\pi_{b_n}$, $a_n$, and $\pi_{a_n}$ in favor of $f_n$ and $\pi_{f_n}$.
At this point we want to emphasize that, unlike the vanishing modes $(b_n,\pi_{b_n})$,
neither $a_n$ nor $\pi_{a_n}$ is equal to zero. Therefore, the reduced
symplectic structure in the gauge-fixed system has a canonical form only in a
suitable new set of variables on the reduced phase space. Let us introduce the following
set:
\begin{eqnarray}\label{eq:gauge_A_new_variables}
\tilde{\alpha}&=&\alpha + \epsilon^2\sum_n\frac{a_n^2}{2}, \\
\tilde{\varphi}&=&\varphi+\epsilon^2\sum_n3a_nf_n, \\
\tilde\pi_\alpha&=&\pi_\alpha,\quad \tilde\pi_\varphi=\pi_\varphi,\\
\tilde f_n&=& f_n,\quad \tilde\pi_{f_n}= \pi_{f_n}-3a_n\pi_\varphi,
\end{eqnarray}
where, after reduction, it is understood that
\begin{eqnarray}\label{eq:gauge_A_pian}
\pi_{a_n}&=&a_n\tilde\pi_\alpha+3f_n\tilde{\pi}_\varphi,\\ \label{eq:gauge_A_an}
a_n&=&3\frac{\tilde\pi_\varphi\tilde{\pi}_{f_n}+(e^{6\tilde\alpha}\tilde m^2\tilde{\varphi}-3\tilde\pi_\alpha\tilde\pi_\varphi)\tilde f_n}{e^{4\tilde\alpha}(n^2-4)},
\end{eqnarray}
at the considered perturbative order.
For simplicity in the notation, we have used $a_n$ in most of the
expressions above, but one must keep in mind that its value is given by
Eq.~\eqref{eq:gauge_A_an}.

The reduced Hamiltonian constraint in this new set of variables, truncated
to the correct perturbative order, is
\begin{eqnarray}\label{eq:reduc_Ham}
\tilde{\mathbb{H}}&=&N_0\bigg(\tilde H_0+\epsilon^2\sum_n\tilde H_2^n\bigg),
\end{eqnarray}
where $\tilde H_0$ has formally the same expression as $H_0$, but now in terms of the new variables
$\tilde{\alpha}$, $\tilde{\varphi}$, $\tilde\pi_\alpha$, and $\tilde\pi_\varphi$
[see Eq.~\eqref{eq:hom_C0}], and
\begin{eqnarray}\label{eq:reduc_Ham1}\nonumber
\tilde H_2^n &=&\frac{e^{-3\tilde \alpha}}{2} \Big( \tilde \pi_{f_n}^2\tilde E^n_{\pi\pi}+\tilde f_n\tilde \pi_{f_n}\tilde E^n_{f\pi}+\tilde f_n^2\tilde E^n_{ff}\Big),\\
\tilde E^n_{\pi\pi}&=&1-\frac{3\tilde\pi_\varphi^2}{e^{4\tilde\alpha}(n^2-4)},\\  \nonumber
\tilde E^n_{f\pi}&=&6\tilde\pi_\varphi\frac{3\tilde\pi_\alpha\tilde\pi_\varphi-e^{6\tilde\alpha}\tilde m^2\tilde{\varphi}}{e^{4\tilde\alpha}(n^2-4)},\\  \nonumber
\tilde E^n_{ff}&=&e^{4\tilde\alpha}(n^2-1)+e^{6\tilde\alpha}\tilde m^2-9\tilde\pi_\varphi^2\\ \nonumber
&-&3\frac{(e^{6\tilde\alpha}\tilde m^2\tilde{\varphi}-3\tilde\pi_\alpha\tilde\pi_\varphi)^2}{e^{4\tilde\alpha}(n^2-4)}.
\end{eqnarray}

Finally, we deal with the modes $n=1$ and $2$. For each of these cases, we
only have two configuration variables, $a_n$ and $f_n$, and their corresponding momenta. Owing
to the presence of the diffeomorphism constraint and the linear scalar
constraint, these modes are completely constrained. A convenient gauge fixing for them
is $a_n=f_n=0$. Then, the constraints $H_{|1}^n=H_{\_\!\_1}^n=0$
imply that their corresponding momenta vanish. These conditions are stable under
the dynamics if $g_n=k_n=0$.

\subsection{Canonical transformation}\label{sec:can_transf}

Now that the system has been reduced, we will change variables on the canonical phase
space to adapt it to the requirements of the uniqueness results provided in
Refs.~\cite{uniq,fmo}, regarding the quantization of fields in non-stationary
scenarios after deparametrization. Specifically, any $SO(4)$-invariant Fock representation that
implements the dynamics unitarily is only compatible with a particular choice of
variables on the phase space of the field, as any genuinely time-dependent linear
canonical transformation of those variables prevents the simultaneous
fulfillment of both properties in another representation~\cite{comment2}. We can
introduce that preferred set of variables by means of a canonical
transformation on the reduced phase space. Let us start with: i)~a scaling of
the field of modes $\tilde f_n$ by the FRW scale factor, and ii)~the inverse
scaling of its momentum, also allowing a suitable momentum shift proportional to the
configuration variable. These modifications can be extended straightforwardly to a
canonical transformation which involves both the homogeneous and the inhomogeneous
sectors. In this way, we arrive at the following variables, which are canonical
at the considered perturbative order:
\begin{eqnarray}\label{eq:canon_trans}\nonumber
\bar \alpha&=&\tilde\alpha+\frac{\epsilon^2}{2}\sum_n \tilde{f}_n^2,\quad
\bar \varphi=\tilde \varphi, \quad \bar\pi_\varphi=\tilde \pi_\varphi,\\ \nonumber
\bar{\pi}_\alpha&=&\tilde\pi_\alpha+\epsilon^2\sum_n \big(\tilde\pi_\alpha\tilde{f}_n^2-\tilde{f}_n\tilde{\pi}_{f_n}\big),\\
\bar f_n&=&e^{\tilde\alpha}\tilde f_n, \quad
\bar \pi_{f_n}=e^{-\tilde\alpha}(\tilde{\pi}_{f_n}-\tilde\pi_\alpha\tilde f_n).
\end{eqnarray}

After the change,
the homogeneous part of the Hamiltonian constraint remains formally the same, though with the old homogeneous variables replaced with the new ones,
while the perturbed
Hamiltonian at order $\epsilon^2$ is
\begin{eqnarray}\label{eq:fundam_Ham}
\bar H_2^n &=&\frac{e^{-\bar \alpha}}{2} \Big( \bar \pi_{f_n}^2\bar E^n_{\pi\pi}+\bar f_n\bar \pi_{f_n}\bar E^n_{f\pi}+\bar f_n^2\bar E^n_{ff}\Big),\\ \nonumber
\bar E^n_{\pi\pi}&=&1-\frac{3 \bar \pi_\varphi^2}{e^{4\bar \alpha}(n^2-4)},\\  \nonumber
\bar E^n_{f\pi}&=& 6 \bar \pi_\varphi\frac{2\bar \pi_\alpha \bar \pi_\varphi-e^{6\bar \alpha}\tilde m^2\bar \varphi}{e^{6\bar \alpha}(n^2-4)},\\  \nonumber
\bar E^n_{ff}&=&n^2-1-
\frac{\bar\pi_\alpha^2+15\bar \pi_\varphi^2-e^{4\bar \alpha}+3e^{6\bar \alpha}\tilde m^2\bar \varphi^2}{2e^{4\bar \alpha}}\\
&+&e^{2\bar \alpha}\tilde m^2-3\frac{(2\bar \pi_\alpha \bar \pi_\varphi-e^{6\bar \alpha}\tilde m^2 \bar \varphi )^2}{e^{8\bar \alpha}(n^2-4)}.
\end{eqnarray}

This description of the reduced phase space, and of its Hamiltonian
constraint, will be our starting point for the hybrid approach, except for a
convenient transformation in the homogeneous sector of the phase space, carried out to reintroduce a natural set of variables for
its polymeric quantization. This transformation leads from $\bar\alpha$, $\bar\pi_\alpha$,
$\bar\varphi$, and $\bar\pi_\varphi$ to variables that are formally similar to $p$, $c$, $\phi$, and
$p_\phi$, namely the variables introduced in Sec.~\ref{sec:hom_class}, though defined now in coexistence with the inhomogeneities.
The change is given again by Eqs. \eqref{changehomo} and \eqref{changehomo2}, replacing the
old variables with their barred counterpart. Besides, we will call $C_0$ and $C_2^n$, respectively, the constraints
$\bar H_0$ and $\bar H_2^n$ expressed in terms of these new variables for the homogeneous sector, and multiplied by a factor of $16\pi G= 12 l_0^3\sigma^2$ in order to adopt the usual conventions employed in LQC.

\section{Hybrid quantization}\label{sec:hybrid_quant}

As we have anticipated, our main aim is to accomplish a complete quantization
of this inflationary model containing inhomogeneities. As a preliminary step, we will
introduce an auxiliary quantum framework --the so-called kinematical Hilbert
space-- to construct a quantum representation of the classical system, and in
particular, of the Hamiltonian constraint. In order to determine the physical
sector, we will look for the states annihilated by this quantum constraint, and build
the physical Hilbert space out of them.

\subsection{Kinematical Hilbert space}\label{sec:hybrid_kin}

Let us consider again the kinematical Hilbert space $\mathcal{H}^{\rm mat}_{\rm kin}\otimes
\mathcal{H}^{\rm grav}_{\rm kin}$ for the homogeneous sector, as explained in
Sec.~\ref{sec:hom_quant}. We recall that, in $\mathcal{H}^{\rm
mat}_{\rm kin}$, the operator $\hat \phi$ acts by multiplication and its
conjugate variable $\hat p_\phi$ is a derivative operator. In the gravitational
sector, the fundamental variables are fluxes and holonomies of
$su(2)$-connections, essentially represented, respectively, by the
multiplicative operator $\hat p$ and the matrix elements of the holonomies,
i.e., $\hat N_{\bar\mu}$ (in the improved dynamics scheme). Both operators have a well
defined action on $ \mathcal{H}^{\rm grav}_{\rm kin}$.

For the perturbations, in Sec.~\ref{sec:can_transf} we arrived at a privileged
description, given by the variables $\bar f_n$ and $\bar\pi_{f_n}$, defined in
Eq.~\eqref{eq:canon_trans}. In this description, the massless representation
permits a unitary quantum implementation of the field dynamics~\cite{fmo}. To
perform a standard Fock quantization, it is hence convenient to rewrite these modes in
terms of the corresponding annihilation and creation-like variables, namely $a_{f_n}$ and their complex conjugate $a_{f_n}^*$, with
\begin{eqnarray}\label{eq:creat-like}
a_{f_n}=\frac{1}{\sqrt{2\omega_n}}(\omega_n \bar f_n+i\bar\pi_{f_n}),
\end{eqnarray}
and $\omega_n^2=n^2-1$.
We promote these variables to quantum operators $\hat a_{f_n}$ and $\hat
a^{\dagger}_{f_n}$, such that $[\hat a_{f_n},\hat
a^{\dagger}_{f_{n'}}]=\delta_{nn'}$. Let us now call $\mathcal S$ the vector space
consisting of finite linear combinations of the $N$-particle states
\begin{eqnarray}\label{eq:nstate}
|\mathcal{N} \rangle= |N_3 , N_4 , ...\rangle, \quad \sum_{n \geq 3}N_n<\infty,
\end{eqnarray}
with $ N_n \in\mathbb{N}$ being the number of particles of the $n$-th mode.
Then, we can construct the inhomogeneous sector of the kinematical Hilbert
space, $\mathcal{F}$, as the completion of $\mathcal S$ with
respect to the inner product $\langle \mathcal{N} |\mathcal{N'}\rangle =
\delta_{\mathcal{N}\mathcal{N}'}$. Obviously, the $N$-particle
states are then an orthonormal basis of the Fock space $\mathcal{F}$.
The total kinematical Hilbert space can thus be taken as
\begin{eqnarray}\label{eq:hilb_kin}
\mathcal{H}^{\rm kin}=\mathcal{H}^{\rm kin}_{\rm grav}\otimes \mathcal{H}^{\rm kin}_{\rm mat}\otimes\mathcal{F}.
\end{eqnarray}
On this Hilbert space, it is understood
that both $\hat a_{f_n}$ and $\hat a^{\dagger}_{f_n}$ act as the identity on the
homogeneous sector.

\subsection{Quantum constraint} \label{sec:hybrid_quan_const}

Let us now concentrate our discussion on the Hamiltonian constraint \eqref{eq:fundam_Ham} after
rewriting it in terms of the variables $p$, $c$, $\phi$, and $p_\phi$, as we have commented above.
To introduce a(n at least symmetric) operator that represents this constraint on $\mathcal{H}^{\rm kin}$, we
admit the natural assumption that the superselection sectors of the homogeneous system (i.e., the sectors $\mathcal{H}^{\varepsilon}_+$, if we are already restricting the analysis to positive $p$) are
preserved by the inclusion of the inhomogeneities, since the latter are regarded as perturbations of the homogeneous setting. We then adopt the following
quantization prescriptions:
\begin{enumerate}
\item Contributions like $\phi p_\phi$ will be promoted to the
operator $(\hat\phi \hat p_\phi+\hat p_\phi \hat\phi )/2$.

\item In any product of a power of the volume $V$ and a non-commuting expression, the former will be evenly distributed around the latter so as to obtain a symmetric combination.

\item Any even power of the form $[(c-l_0)p]^{2k}$, with $k\in\mathbb{Z}$, will be represented
by $\widehat{\Theta}^e_{(k)}=[\widehat{\Omega}^{2}]^{k}$, constructed using the positive operator
$\widehat{\Omega}^{2}$ [see Eq.~\eqref{eq:quantumomega}] and the spectral theorem to define its $k$-th power~\cite{reedsi,comment3}.

\item In the case of odd powers of the form $[(c-l_0)p]^{2k+1}$, the prescription
will be
\begin{eqnarray}
[(c-l_0)p]^{2k+1}\to \widehat{\Theta}^o_{(k)}=|\widehat{\Omega}|^{k} \widehat{\Lambda}|\widehat{\Omega}|^{k},
\end{eqnarray}
where $|\widehat{\Omega}|=\sqrt{\widehat{\Omega}^2}$ and
\begin{eqnarray}\label{eq:quantumlambda}
\widehat{\Lambda}&=&
\frac{\hat{V}^{1/2}}{8i\sqrt{\Delta}}
\Big[\widehat{{\rm sgn}(v)} \sum_{r=+1,-1}\left(r e^{-ir\hat{\bar{\mu}}l_0}\hat{N}_{4r\bar{\mu}}e^{-ir\hat{\bar{\mu}}l_0}\right)
\nonumber\\  &+&\sum_{r=+1,-1}\left(r e^{-ir\hat{\bar{\mu}}l_0}\hat{N}_{4r\bar{\mu}}e^{-ir\hat{\bar{\mu}}l_0}\right) \widehat{{\rm sgn}(v)}\Big]\hat{V}^{1/2}.
\end{eqnarray}
\end{enumerate}

One can straightforwardly see that $\widehat{\Lambda}$ is a difference operator
which only relates states with support in semilattices of step 4 for
$v$. Therefore, any power of $(c-l_0)p$ will be promoted indeed to an operator
that preserves the superselection sectors in this
homogeneous volume (see Sec.~\ref{sec:hom_quant}).

Using these assignments, we arrive at a quantum constraint of the form
\begin{eqnarray}\label{eq:quantum_constA}
\hat C&=&\hat C_0+\epsilon^2\sum_n\hat C_2^n,
\end{eqnarray}
where $\hat C_0$ is the homogeneous constraint defined in
Eq.~\eqref{eq:hom_quantum_const}, and
\begin{eqnarray}\label{eq:inh_quan_const}
\hat C^n_2&=&6l_0^4\sigma^2 \widehat{\left[\frac{1}{V}\right]}^{1/6}\bigg[\hat N^n\bigg(2\omega_n+\frac{1}{\omega_n}\hat F^n_{-}\bigg)\\ \nonumber
&+&\frac{1}{2\omega_n}\bigg(\hat X_+^n \hat F^n_{+}+\frac{3i\omega_n \sigma^2}{\omega_n^2-3}\hat X_-^n \hat G\bigg)
\bigg]\widehat{\left[\frac{1}{V}\right]}^{1/6}.
\end{eqnarray}
Here
\begin{eqnarray}\label{eq:CoeffA}
\hat N^n&=&\hat a^\dagger_{f_n} \hat a_{f_n}, \quad \hat X_\pm^n= (\hat a^\dagger_{f_n})^2\pm \hat a_{f_n}^2,\\ \nonumber
\hat F_\pm^n&=&-
\frac{\sigma^2l_0}{2}\widehat{\left[\frac{1}{V}\right]}^{2/3}\Bigg(15\hat p_\phi^2+3m^2\hat V^2\hat\phi^2+\frac{\widehat{\Theta}^e_{(1)}}{\gamma^2l_0^3\sigma^2}\Bigg)\widehat{\left[\frac{1}{V}\right]}^{2/3}\\ \nonumber
&-&\frac{3}{n^2-4}\frac{\sigma^2}{l_0}\widehat{\left[\frac{1}{V}\right]}^{4/3}\bigg(\frac{2}{\gamma}\hat p_\phi\widehat{\Theta}^o_{(0)}
+m^2\hat V^2 \hat \phi \bigg)^2\widehat{\left[\frac{1}{V}\right]}^{4/3}\nonumber\\
&+&\frac{1}{2}+\frac{m^2}{l_0^2}\hat p
\pm 3 \sigma^2l_0 \frac{n^2-1}{n^2-4}\hat p_\phi^2\widehat{\left[\frac{1}{V}\right]}^{4/3},\\
\hat G &=& -\widehat{\left[\frac{1}{V}\right]}\bigg[m^2\hat p^3(\hat\phi \hat p_\phi+\hat p_\phi \hat\phi)+\frac{4}{\gamma}\hat p^2_\phi\widehat{\Theta}^o_{(0)} \bigg]\widehat{\left[\frac{1}{V}\right]}.
\end{eqnarray}
At this stage, it is worth noticing that any possible factor ordering ambiguity affecting $\hat C^n_2$, resulting from prescriptions other than ours for its operator representation, produces only subleading quantum geometry corrections to the (already) perturbative terms in the total constraint.

As in the absence of inhomogeneities, the constraint $\hat C$ annihilates the state $|v=0\rangle$ (times any state in $\mathcal{H}^{\rm kin}_{\rm mat}\otimes\mathcal{F}$), and leaves invariant its orthogonal complement, $\widetilde{\mathcal{H}}^{\rm kin}$, complement to which we can hence restrict all considerations in the following, removing the state analog of the cosmological (homogeneous) singularity. Besides, as anticipated,
the operator  $\hat C$ preserves the sectors of superselection in the homogeneous volume, namely the (positive) semilattices of step 4 in the label $v$, since all the basic operators from which $\hat C$ is constructed preserve those sectors in fact. In other words, as in the analysis of the homogeneous system, the action on $v$ of all the operators considered in this article preserve the subspaces $\mathcal{H}^{\varepsilon}_+$ of states with support in the semilattices $\mathcal{L}_{\varepsilon}^+$ (limiting again the discussion to positive values of $v$ for simplicity).

\subsection{Physical Hilbert space}\label{sec:hybrid_Phys}

In order to complete our quantization, we still have to characterize the space of solutions to the constraint
$\hat C$ and provide it with the structure of a Hilbert space. In accordance with our perturbative approach, we will assume that
the solutions can be expanded in a perturbative series and truncate them in the form
\begin{eqnarray}\label{eq:sol_pert_exp}
(\psi|=(\psi|^{(0)}+\epsilon^2(\psi|^{(2)}.
\end{eqnarray}
In turn, each term can be expanded employing the basis that we have introduced for the total kinematical Hilbert space of the system:
\begin{equation}
(\psi|^{(k)}=\sum_{\mathcal{N}}\sum_{v\in\mathcal{L}_{\varepsilon}}\int d\phi\langle \mathcal{N}|\otimes\langle v|\otimes\langle\phi |\psi^{(k)}(\mathcal{N},v,\phi).
\end{equation}

By consistency, the Hamiltonian
constraint
$(\psi|\hat C^\dagger$ must vanish order by order, up to the level of our approximation in the perturbative expansion. Here, the dagger denotes again the adjoint. The zeroth-order contribution to the constraint yields $(\psi|^{(0)}\hat
C_0^\dagger=0$. But this condition is just the constraint studied in
Sec.~\ref{sec:hom_quant} for the homogeneous sector. Recall that we have shown that the corresponding solutions
are completely characterized by the initial data on the minimum volume section $v=\varepsilon$. In addition,
note that, at this zeroth order, the information that $(\psi|^{(0)}$ codifies about the
inhomogeneities does not change in the evolution in $v$ that the constraint $\hat
C_0^\dagger$ dictates.

The next term in the perturbation expansion gives
\begin{eqnarray}\label{eq:eq_pert_count_solut}
(\psi|^{(2)}\hat C_0^\dagger+(\psi|^{(0)} \Big(\sum_n\hat C_2^n\Big)^\dagger &=&0.
\end{eqnarray}
This relation tells us that $\psi^{(2)}$ must satisfy a difference equation similar to that for $(\psi|^{(0)}$, but
now with a source term which accounts for the interaction of the
inhomogeneities with the ``background'' state $(\psi|^{(0)}$. Again, this can be
interpreted as an evolution equation in the internal time $v$, in which any
solution emerges out of the initial section without the need for any boundary
condition around it. Consequently, if the initial data $\psi^{(2)}|_{v=\varepsilon}$ is
provided, one can straightforwardly determine
$\psi^{(2)}|_{v=\varepsilon+4}$ once $(\psi|^{(0)}$ is supplied: one only needs to solve a (well-posed) linear difference equation in
$v$. By the same arguments, following an iterative process we can find the value of
the solution on any other section $v=\varepsilon + 4n$, $\forall n \in
\mathbb{N}$. In conclusion, any solution $(\psi|$ to the constraint can be
determined if the initial data on the minimum volume section
($v=\varepsilon$) are given. Although the construction of the solutions is formal,
in general, we will see that this suffices to determine a physical Hilbert space which retains
all the true degrees of freedom of the system.

To do that, we will proceed in a similar way as we did for the homogeneous part of the model in
Sec.~\ref{sec:hom_quant}. Again, the important point is the fact that we can identify solutions with
their initial data on the minimum volume section. Therefore, we just need to equip that
space of initial data with a suitable inner product in order to construct the
physical Hilbert space (see Ref.~\cite{gowdy1} for additional discussion). This inner product can be fixed, for instance,
by demanding reality conditions on a complete set of observables \cite{reacon}. Implementing this approach, the
physical Hilbert space that we obtain is then
\begin{eqnarray}\label{eq:phys_hilb}
\mathcal{H}^{\rm phys}= \mathcal{H}^{\rm kin}_{\rm mat}\otimes\mathcal{F}.
\end{eqnarray}

\section{An alternate gauge fixing}\label{sec:gauge_B}

In addition to the previous analysis, we will consider now an alternate gauge
fixing in which the spatial metric looks homogeneous (and hence also the
spatial curvature). We will carry out the reduction and quantization of the system in a
similar way to what we did in the gauge discussed so far.

Let us start again with the constraint given in Eqs.~\eqref{eq:scala_const2},
\eqref{eq:scala_const1}, and \eqref{eq:diffeo_const}. We will impose the
conditions $a_n=0$ and $b_n=0$, which are second class with the linear constraints.
The stability of this gauge under the evolution can be ensured by fixing $g_n$
and $k_n$ as certain linear functions of the momenta  $\pi_{a_n}$ and
$\pi_{b_n}$, whereas these momenta can be written as functions of the variables
$f_n$ and $\pi_{f_n}$ --and the homogeneous variables-- once the constraints
$H_{|1}^n=0$ and $H_{\_\!\_1}^n=0$ are satisfied exactly.

Specifically, requiring that $\{b_n,\mathbb{H}\}=0$, we
arrive again at the very same condition given in Eq.~\eqref{eq:gauge_A_bn}, while (together with this last equation)
the consistency demand $\{a_n,\mathbb{H}\}=0$ leads to
\begin{eqnarray}\label{gf1}
g_n&=&\frac{1}{\pi_\alpha}\left(\frac{n^2-1}{n^2-4}\pi_{b_n}-\pi_{a_n}\right),
\end{eqnarray}
where we are using the following relations, obtained in the reduction process:
\begin{eqnarray}\label{eq:gauge_condB}
\pi_{a_n}&=&\frac{\pi_\varphi}{\pi_\alpha}\pi_{f_n}+\frac{\tilde{m}^2e^{6\alpha}\varphi}{\pi_\alpha}f_n,\\\nonumber \pi_{b_n}&=&\frac{\pi_\varphi}{\pi_\alpha}\pi_{f_n}+\bigg(\frac{\tilde{m}^2e^{6\alpha}\varphi}{\pi_\alpha}-3\pi_\varphi\bigg)f_n.
\end{eqnarray}
After introducing these last expressions in Eq.~\eqref{eq:scala_const2}, we
arrive at a Hamiltonian constraint of the same type as in Eq.~\eqref{eq:reduc_Ham},
but with the new coefficients
\begin{eqnarray}\label{eq:coeff_reduc_HamB}
E^n_{\pi\pi}&=&1+\frac{3}{n^2-4}\frac{\pi_\varphi^2}{\pi_\alpha^2},\nonumber\\ \nonumber
E^n_{f\pi}&=&-6\frac{\pi_\varphi^2}{\pi_{\alpha}} +\frac{3}{n^2-4}\bigg(\frac{2
e^{6\alpha}\tilde{m}^2 \varphi \pi_\varphi}{\pi_\alpha^2}-\frac{6\pi_\varphi^2}{\pi_\alpha}\bigg),\\
\nonumber E^n_{ff}&=&e^{4\alpha}(n^2-1)+e^{6\alpha}\tilde{m}^2+9\pi_\varphi^2-\frac{6
e^{6\alpha}\tilde{m}^2 \varphi \pi_\varphi}{\pi_\alpha}\,\,\,\,\,\,\,\,\,\,\,\,\\
&+&\frac{3}{n^2-4}\bigg(3\pi_\varphi-\frac{e^{6\alpha}\tilde{m}^2 \varphi }{\pi_\alpha}\bigg)^2.
\end{eqnarray}

It is evident from all these equations that the introduced gauge fixing
is always well posed except on the section of phase space corresponding to
vanishing momentum $\pi_\alpha$. We will see later that even the potential
problems posed on that section are eluded in our quantization, owing to our regularization
prescription and the fact that the kernel of the quantum counterpart of $\pi_\alpha$
belongs to the continuum spectrum, so that its corresponding operator can be inverted
via the spectral theorem \cite{reedsi}.

We will now
introduce a scaling of the configuration variables $f_n$ by the FRW scale
factor, extended to a complete canonical transformation, so that the new canonical pair of fields that describe the matter perturbation admit a Fock quantization with the good properties explained in Sec.~\ref{sec:can_transf}. To this end, in addition to the scaling of the $f_n$'s, we
perform the inverse scaling of the conjugate momenta, to which we also add a
term linear in their corresponding configuration field variables in order to
ensure that the cross-term coefficients $E^n_{f\pi}$ have a subdominant contribution to the matter field dynamics in the
large $n$ limit. Of course, we must transform the homogeneous variables as well,
so that the entire transformation on phase space is canonical at the considered
perturbative order. Explicitly, the canonical change is given by
\begin{eqnarray}\label{eq:canon_transB}\bar \alpha&=&\alpha+\frac{1}{2}\bigg(1-3\frac{\pi_\varphi^2}{\pi_\alpha^2}\bigg)\epsilon^2\sum_n f_n^2, \nonumber \\
\bar{\pi}_\alpha&=&\pi_\alpha+\epsilon^2\sum_n \bigg[\bigg(3\frac{\pi_\varphi^2}{\pi_\alpha}+\pi_\alpha\bigg) f_n^2-f_n\pi_{f_n}\bigg],\nonumber\\
\bar \varphi&=& \varphi+3\frac{\pi_\varphi}{\pi_\alpha}\epsilon^2\sum_n f_n^2, \quad \quad \bar\pi_\varphi=\pi_\varphi, \nonumber\\
\bar f_n&=&e^{\alpha} f_n, \quad
\bar \pi_{f_n}=e^{-\alpha}\bigg[\pi_{f_n}-\bigg(3\frac{\pi_\varphi^2}{\pi_\alpha}+\pi_\alpha\bigg) f_n\bigg].\,\,\,\,\,\,\,\,
\end{eqnarray}

Under this transformation, the homogeneous part of the Hamiltonian constraint~\eqref{eq:hom_C0} is kept formally the same, though with the old homogeneous variables replaced with their new counterparts, whereas, in terms of the new inhomogeneous variables, the contributions quadratic in the perturbations have again the general form given by Eq.~\eqref{eq:fundam_Ham}, but now with coefficients
\begin{eqnarray}\label{eq:fundam_HamB}\nonumber
\bar E^n_{\pi\pi}&=&1+\frac{3}{n^2-4}\frac{\bar \pi_\varphi^2}{\bar\pi_\alpha^2},\\ \nonumber
\bar E^n_{f\pi}&=&\frac{6}{e^{2\bar\alpha}(n^2-4)}\bigg(\frac{e^{6\bar\alpha}\tilde{m}^2 \bar\varphi \bar \pi_\varphi}{\bar\pi_\alpha^2}-\frac{2\bar \pi_\varphi^2}{\bar \pi_\alpha}+\frac{3\bar \pi_\varphi^4}{\bar\pi_\alpha^3}\bigg),\\  \nonumber
\bar E^n_{ff}&=&n^2-1-
\frac{1}{2e^{4\bar\alpha}}\bigg(\bar\pi_\alpha^2-30\bar \pi_\varphi^2+\frac{27\bar \pi_\varphi^4}{\bar\pi_\alpha^2}-e^{4\bar\alpha}\bigg)\\ \nonumber
&+&e^{2\bar \alpha}\tilde{m}^2-\frac{3e^{2\bar\alpha}}{2}\tilde{m}^2 \bar\varphi\bigg[\frac{8 \bar \pi_\varphi }{\bar \pi_\alpha}-\bar\varphi\bigg(\frac{3 \bar \pi_\varphi^2 }{\bar\pi_\alpha^2}-1\bigg)\bigg]-\frac{3\bar \pi_\varphi^2}{2\bar\pi_\alpha^2}\\
&+&\frac{3}{e^{4\bar\alpha}(n^2-4)}\bigg(\frac{e^{6\bar\alpha}\tilde{m}^2 \bar\varphi }{\bar\pi_\alpha}-2\bar \pi_\varphi+\frac{3\bar \pi_\varphi^3}{\bar\pi_\alpha^2}\bigg)^2.
\end{eqnarray}

Finally, let us consider the variables $c$, $p$, $\phi$, and $p_\phi$
applied in the quantization of the homogeneous system, already described in
Sec.~\ref{sec:hom_sys}. Again, they are given by Eqs.~\eqref{changehomo} and \eqref{changehomo2}, replacing the
old variables with their barred counterparts. Once more, we can
combine a polymeric quantization of this homogeneous sector together with a standard Fock quantization of the inhomogeneities,
providing a kinematical arena for the quantum treatment
of the system. Following the quantization prescription of
Sec.~\ref{sec:hybrid_quan_const}, the reduced Hamiltonian constraint, with
coefficients given by Eq.~\eqref{eq:fundam_HamB}, can be promoted to an adequate
operator, like the one introduced in Eq.~\eqref{eq:quantum_constA} and with a
similar contribution~\eqref{eq:inh_quan_const} of the inhomogeneities, but now
with the operators
\begin{widetext}
\begin{eqnarray}\label{eq:CoeffB}\nonumber
\hat F_\pm^n&=&-
\frac{3\sigma^2l_0}{2}\bigg[m^2|\hat p|^{1/2}\left(-4\gamma(\hat\phi\hat p_\phi+\hat p_\phi \hat\phi) \widehat{\Theta}^o_{(-1)}-\frac34\gamma^2\sigma^2l_0^3(\hat\phi\hat p_\phi+\hat p_\phi\hat\phi)^2\widehat{\Theta}^e_{(-1)}+\hat\phi^2\right)|\hat p|^{1/2}+\gamma^2l_0^2\hat p_\phi^2\widehat{\Theta}^e_{(-1)}\bigg]\\ \nonumber
&+&
\frac{\sigma^2l_0}{2}\widehat{\left[\frac{1}{V}\right]}^{2/3}\Bigg(30\hat p_\phi^2-27\gamma^2\sigma^2l_0^3\hat p_\phi^4\widehat{\Theta}^e_{(-1)}-\frac{\widehat{\Theta}^e_{(1)}}{\gamma^2l_0^3\sigma^2}\Bigg)\widehat{\left[\frac{1}{V}\right]}^{2/3}
+
\frac{3\sigma^2l_0}{n^2-4}\widehat{\left[\frac{1}{V}\right]}^{2/3}\bigg(2\hat p_\phi +\gamma m^2|\hat p|^{3/2}\widehat{\Theta}^o_{(-1)}|\hat p|^{3/2}\hat \phi  \\
&-& 3\gamma^2\sigma^2 l_0^3 \hat p^3_\phi\widehat{\Theta}^e_{(-1)}\bigg)^2\widehat{\left[\frac{1}{V}\right]}^{2/3}+\frac{1}{2}+\frac{m^2}{l_0^2}\hat p \mp 3\sigma^2l_0^3 \gamma^2 \frac{n^2-1}{n^2-4}\hat p^2_\phi\widehat{\Theta}^e_{(-1)},\\  \nonumber
\hat G &=& 2\gamma l_0^2 \widehat{\left[\frac{1}{V}\right]}^{1/3}\bigg[\frac{\gamma m^2}{2}(\hat\phi\hat p_\phi+\hat p_\phi \hat\phi)|\hat p|^{3/2}\widehat{\Theta}^e_{(-1)}|\hat p|^{3/2}+2\hat p_\phi^2\widehat{\Theta}^o_{(-1)}-3\gamma^2\sigma^2 l_0^3\hat p_\phi^4 |\hat p|^{1/2}\widehat{\Theta}^o_{(-2)}|\hat p|^{1/2} \bigg]\widehat{\left[\frac{1}{V}\right]}^{1/3}.
\end{eqnarray}
\end{widetext}
This Hamiltonian constraint also decouples the zero volume state
(tensor product any state on the matter field, both for its homogeneous part and its inhomogeneities), and is a
combination of operators that preserve the superselection sectors of the
homogeneous sector: semilattices of step 4 in the label $v$. On the other hand, notice the appearance of negative powers of $\hat \Omega^2$ through the terms $\widehat{\Theta}^o_{(-1)}$, $\widehat{\Theta}^o_{(-2)}$, and $\widehat{\Theta}^e_{(-1)}$. This inverse powers are well defined via the spectral decomposition of $\hat \Omega^2$, since the kernel of this operator is in its continuum spectrum \cite{comment3}.

Physical states are annihilated by the operator $\hat C$.
We will assume the same perturbative form for them as in
Eq.~\eqref{eq:sol_pert_exp} (at the order studied in our approximation). The part $(\psi|^{(0)}$ of the
solutions represents again a background-like state. Following the analysis
carried out in Sec.~\ref{sec:hom_sys}, any such solution (at zeroth perturbative order) is characterized by
its initial data on the minimum volume section. Once that background
solution is determined, one can evaluate $(\psi|^{(2)}$, which takes into account
the interplay between the inhomogeneities and the homogeneous background.
In Sec.~\ref{sec:hybrid_Phys} we saw that
this part of the solution can be completely determined from its initial
data at $v=\varepsilon$ once $(\psi|^{(0)}$ is known, at
least formally. In total, we conclude that the space of solutions to the constraint $\hat C$ can be identified
with the space of data on the minimum volume section. Finally, we can select a
convenient inner product on this space, e.g., by
choosing a complete set of observables and demanding that the complex conjugation relations between their elements
become adjointness relations between their corresponding operators. In this way, we
reach the same conclusion as in Sec.~\ref{sec:hybrid_Phys}: the physical
Hilbert space can be taken unitarily equivalent to the Hilbert space~\eqref{eq:phys_hilb}.

\section{Gauge-invariant formulation: A unitary map}\label{sec:bardeen}

Gauge-invariant quantities provide a physically
meaningful description of cosmological perturbations~\cite{bardeen}.
Such quantities are usually employed to describe the physics in a
consistent manner, independent of the identification of the spacetime
and its matter content when transformations under diffeomorphisms are allowed,
and insensitive to the specification of a particular gauge. In order to
consolidate our proposal and show the robustness in this respect, in this
section we will establish the correspondence between our fundamental variables $\bar f_n$ and $\bar\pi_{f_n}$ and
gauge-invariant quantities, thus reformulating our quantum description in terms of the latter.
Furthermore, we will see that both formulations (namely, the original one in terms of the matter field perturbations
and the new one in terms of gauge invariants) are related by a canonical transformation which
can be implemented quantum mechanically as a unitary transformation, at least, after deparametrization of the theory
(e.g., in the regime of quantum field theory in a curved background spacetime).

Let us consider a transformation of the general type
$x^{\prime\mu}=x^\mu+\epsilon\xi^\mu$, where $x^\mu$ is a spacetime point,
$\xi^\mu$ is an arbitrary vector, and $x^{\prime\mu}$ is the transformed point
(again, $\epsilon$ is the perturbative parameter introduced in
Sec.~\ref{sec:inhom_class_pert}). We express the covariant counterpart of $\xi^\mu$
using (hyper\nobreakdash-)spherical harmonics, obtaining the mode expansion
\begin{eqnarray}\label{eq:xi_param}
\xi_0&=&\sigma^2 N_0 \sqrt{2}\pi\sum_n\xi_0^nQ^n,\\
\xi_a&=&\sigma^2 e^{\alpha}
\sqrt{2}\pi\sum_n\xi^nP_a^n.\label{eq:xi_param2}
\end{eqnarray}

On the other hand, instead of following
the procedure presented in Ref.~\cite{scalar1} to obtain
gauge-invariant canonical pairs, here we will rather start from the standard
formalism of Ref.~\cite{bardeen}, where several gauge-invariant perturbations
are defined. In particular, we will consider
the gauge-invariant quantities $\Phi_n^A$, $\Phi_n^B$, $\mathcal{E}^m_n$, and
$v^s_n$ given by Eqs.~\eqref{eq:gauge_inv_phiA}-\eqref{eq:gauge_inv_vs} of Appendix~\ref{appendixC}.
The first couple of invariants is constructed purely out of geometrical quantities. The second one
contains also matter perturbations, and we will refer to them as the
energy density and the velocity gauge-invariant perturbations, respectively.
Any other gauge-invariant variable can be expressed as a linear combination of these
four ones. This is the case of the Mukhanov-Sasaki variable (see
Appendix~\ref{appendixD}), which is a gauge-invariant perturbation {\emph{well adapted
to flat scenarios}}, where it satisfies the equation of motion of a scalar field
propagating in a static spacetime but with a time-dependent potential (see,
e.g.,  Ref.~\cite{scalar1} for a recent discussion). However, for the matter
content considered in this article, the gauge-invariant energy $\mathcal{E}^m_n$
(up to a suitable scaling) is more convenient inasmuch as it satisfies the same type of equation
but independently of the spatial topology, a fact that makes it privileged in comparison
with other invariants.

In fact, one can check that the
combination of Eqs.~\eqref{eq:gauge_inv_rel1} and \eqref{eq:gauge_inv_rel2} with
\eqref{eq:gauge_inv_rel3} and \eqref{eq:gauge_inv_rel4} yields a system of
first order differential equations for the quantities $\mathcal{E}^m_n$ and
$v^s_n$ which resembles the first order equations of a canonical pair of
variables describing an oscillator with time-dependent frequency.
Taking these considerations as our starting point, we define the
following variables
\begin{eqnarray}\label{eq:gauge-inv_canon}\nonumber
\Psi_n&=&\frac{e^{5\alpha}}{\pi_\varphi \sqrt{n^2-4}}E_0\mathcal{E}^m_n,
\\ \Pi_{\Psi_n}&=&-\frac{\sqrt{n^2-4}}{\sqrt{n^2-1}}\frac{\pi_\varphi}{e^{\alpha}} v_n^s+
\bigg(\frac{e^{6\alpha}\tilde{m}^2\varphi}{\pi_{\varphi}}-2 \pi_{\alpha}\bigg)\nonumber\\
&\times& \frac{e^{3\alpha}}{\pi_\varphi \sqrt{n^2-4}}E_0\mathcal{E}^m_n.
\end{eqnarray}
Using Hamilton's equations for $\pi_\varphi$
and $\alpha$, one can actually see that the two introduced variables are related dynamically by
\begin{eqnarray}\label{eq:gauge-inv_canon_eqs1}
\dot\Psi_n=\Pi_{\Psi_n},
\end{eqnarray}
where the dot stands for the derivative with respect to the conformal time (corresponding to $N_0=e^\alpha$).
With the above equation of motion and the corresponding one for $\dot\Pi_{\Psi_n}$,
one obtains a second order differential equation for $\Psi_n$ that corresponds
exactly to that of the modes of a scalar field propagating in a static spacetime with a
time-dependent quadratic potential.

In the rest of this section, we will show that, for each of the two gauge fixations
considered in our discussion, the pair of gauge invariants defined in
Eq.~\eqref{eq:gauge-inv_canon} are related with the fundamental fields $\bar
f_n$ and $\bar\pi_{f_n}$ by means of a canonical transformation.
Moreover, we will also prove that
such a canonical transformation is implementable as a unitary transformation
in the corresponding (deparametrized) quantum theory.

\subsection{Longitudinal gauge $b_n=\Pi_{n}=0$}\label{sec:unitarity_gaugeA}

In Sec.~\ref{sec:inhomog_class} we analyzed this gauge and introduced a canonical pair of
variables $\bar f_n$ and $\bar \pi_{f_n}$, given in Eq.~\eqref{eq:canon_trans},
to describe the inhomogeneities after the corresponding
reduction of the system.
If we now substitute these variables in expressions~\eqref{eq:Em_matter_field} and
\eqref{eq:vs_matter_field}, and then in Eq.~\eqref{eq:gauge-inv_canon},
we arrive at the following relations:
\begin{eqnarray}\label{eq:Pisn_gaugeA}
\Psi_n&=&\frac{1}{\sqrt{n^2-4}}(\bar\pi_{f_n}+\chi_A\bar f_n),\\ \nonumber
\Pi_{\Psi_n}&=&\frac{\chi_A}{\sqrt{n^2-4}}(\bar\pi_{f_n}+\chi_A\bar f_n)-\sqrt{n^2-4}\bar f_n,
\end{eqnarray}
where
\begin{eqnarray}\label{eq:chiA}
\chi_A = e^{4\alpha}\tilde{m}^2\frac{\varphi}{\pi_\varphi} -2\frac{\pi_\alpha}{e^{2\alpha}}
\end{eqnarray}
only depends on the homogeneous variables. Notice also that, at the perturbative order of our approximation,
the substitution in the above relations of the homogeneous variables $\alpha$, $\pi_\alpha$,
$\varphi$, and $\pi_\varphi$ by their barred counterparts appearing in Eq.~\eqref{eq:canon_trans} is
totally irrelevant.

The change \eqref{eq:Pisn_gaugeA} can be regarded as
a Bogoliubov transformation once we consider the creation and annihilation-like
variables associated with each pair of variables, given by
Eq.~\eqref{eq:creat-like} and
\begin{eqnarray}\label{eq:gauge-inv_annih_var}
b^{\Psi}_{n}=\frac{1}{\sqrt{2\omega_n}}(\omega_n\Psi_n+i\Pi_{\Psi_n}),
\end{eqnarray}
together with the complex conjugate. A simple calculation allows us to compute the coefficients associated with the
antilinear part of this Bogoliubov transformation:
\begin{eqnarray}\label{eq:beta_gauge_A}
\beta_n&=&\frac{i}{2}\frac{\chi^2_A+3}{\sqrt{n^2-1}\sqrt{n^2-4}}.
\end{eqnarray}
In order to determine whether the transformation admits or not a unitary implementation, the necessary and
sufficient condition is just that
\begin{eqnarray}\label{eq:unitarity_condition}
\sum_{nlm} |\beta_n|^2=\sum_n d_n|\beta_n|^2<\infty,
\end{eqnarray}
where $d_n$ is a degeneracy factor accounting for the dimension of each
eigenspace of the Laplace-Beltrami operator. Explicitly, we have on the three-sphere that $d_n=n^2$.
As a result, the canonical transformation between our variables
$\bar f_n$ and $\bar\pi_{f_n}$, and the gauge-invariant ones, $\Psi_n$ and
$\Pi_{\Psi_n}$, turns out to be implementable as a unitary one at the quantum level, since the
coefficients $\beta_n$  are clearly square summable, including their degeneracy. Therefore, the two
considered physical
descriptions are completely equivalent.

\subsection{Gauge $a_n=b_n=0$}\label{sec:unitarity_gaugeB}

Let us now consider the gauge fixing carried out in Sec.~\ref{sec:gauge_B}.
Again, one can find the relation in the reduced system between the
canonical pair $\bar f_n$ and $\bar\pi_{f_n}$
and the gauge invariants $\Psi_n$ and $\Pi_{\Psi_n}$. For this, one can
first obtain the
expressions of $\mathcal{E}_n^m$ and $v^s_n$ in terms of $\bar f_n$ and
$\bar\pi_{f_n}$ (in the gauge under consideration) and then use
Eq.~\eqref{eq:gauge-inv_canon}. A simple calculation yields an expression
similar to Eq.~\eqref{eq:Pisn_gaugeA}, but substituting $\chi_A$ by
\begin{eqnarray}
\chi_B = e^{4\alpha}\tilde{m}^2\frac{\varphi}{\pi_\varphi}-2\frac{\pi_\alpha}{e^{2\alpha}}+3\frac{\pi_\varphi^2}{e^{2\alpha} \pi_\alpha}.
\end{eqnarray}
We can then compute the corresponding $\beta$-coefficients of the Bogoliubov
transformation relating the creation and annihilation-like variables associated
with the variables $\bar f_n$ and $\bar\pi_{f_n}$, on the one hand, and the gauge-invariant
variables, on the other hand. The result is the same as in Eq.~\eqref{eq:beta_gauge_A}, but now with
$\chi_B$ replacing $\chi_A$. Following the arguments of
Sec.~\ref{sec:unitarity_gaugeA}, we can easily check the unitary implementability of this
Bogoliubov transformation, since condition~\eqref{eq:unitarity_condition} holds.

\section{Summary and conclusions}\label{sec:conclusions}

In this article, we have presented the quantization of an inflationary universe
with small inhomogeneities. The matter content is described by a minimally
coupled, massive scalar field. First, we have studied the corresponding
symmetry-reduced homogeneous and isotropic system in Sec.~\ref{sec:hom_sys}. In order to
employ a polymeric quantization~\cite{apsv} for the geometry degrees of
freedom, its classical phase space has been parametrized by a densitized triad
and an Ashtekar-Barbero connection. As for the (homogeneous) matter content,
we have applied a standard Schr\"odinger quantization. In this kinematical arena,
we have introduced an operator representation for the Hamiltonian constraint.
The geometric part of this quantum constraint is a difference operator
that only relates states with support in semilattices with
points separated by constant steps of four units. Additionally, these
semilattices can be characterized by a continuous, non-vanishing
parameter $\varepsilon\in(0,4]$, which is proportional to the minimum physical
volume allowed in the quantum theory on that semilattice.
Moreover, we have argued that every sector
of the kinematical Hilbert space with support on any of those semilattices is
superselected. As any physical state belongs to the kernel of the constraint,
one can see that the restriction to a particular superselection sector allows us
to entirely determine any solution if its initial data on the minimum volume
section are provided. The final step to quantize the system to completion, i.e.,
to construct the physical Hilbert space, is to equip the space of solutions with
an adequate inner product. Based on the identification of the space of solutions
with the space of initial data for minimum volume, we have picked up this inner
product by selecting a complete set of observables and imposing reality conditions
on them \cite{reacon}, namely, by
requiring that the complex conjugation relations in this set become adjointness
relations between the operators which represent the observables.
Our proposal allows us not only to
formalize the quantization of the system, but also to proceed in the analysis
of its quantum dynamics. The extension of the systematic procedures commonly
used in LQC to systems without an explicit separation of the matter content from
the geometry is not obvious, yet our proposal sheds new insights to confront
also those more general situations. Our preliminary numerical analysis shows
that, with an appropriate choice of initial data, it is possible to recover
physical states that represent an expanding universe preceded by a contracting
one with a well-defined quantum bounce connecting both of them, even for a
massive scalar field. Those results are very promising and a more careful study
will be the subject of future research.

On the other hand, to progress in the applicability of LQC to the analysis
of the Early Universe and construct a quantum framework to study
the development of small inhomogeneities in inflationary scenarios,
we have introduced local degrees of freedom in our model
by means of perturbation theory around the homogeneous and isotropic solutions. We have
concentrated our efforts in the understanding of scalar perturbations, because
of the fundamental role that they play in present observational cosmology. Additionally,
from the technical point of view, scalar modes are more involved than vector or
tensor ones, since they incorporate both physical and gauge degrees of freedom and
their discrimination is not trivial. In order to distinguish the true
physical degrees of freedom, we have carried out two different partial gauge
fixings. The first one is the so-called longitudinal gauge, commonly used in
standard cosmology, and the second one is the natural gauge fixing in which the
spatial metric is purely homogeneous (i.e., all the inhomogeneities of the metric are
encoded in the perturbations of the lapse and shift functions). In both cases,
we reach a description of the inhomogeneities of the corresponding reduced system in terms of
the matter perturbations. We have introduced a canonical transformation in each
case, with an eye on the standard Fock quantization of the inhomogeneities, quantization
that we have carried out later on. Such a new choice of canonical fields is motivated by the recent
uniqueness results regarding the quantization of linear fields (generically) in non-stationary
spacetimes \cite{uniq}, where the choice of a Fock quantum theory with a vacuum
state invariant under the spatial isometries and unitary dynamics
selects privileged canonical field variables for the description of the system,
together with a specific quantum representation for such field variables. We have then
adopted this Fock representation
for the inhomogeneities and combined it with the polymeric description initially introduced in
Sec.~\ref{sec:hom_quant}.

Furthermore, we have presented a quantization
prescription for the Hamiltonian constraint in Sec.~\ref{sec:hybrid_quan_const}.
The corresponding quantum operator respects the superselection sectors of the
homogeneous setting. Therefore, any state which is a solution to the constraint
has support in semilattices of constant
step in the physical volume, starting from a non-vanishing minimum value of it.
We have been able to prove that, assuming that the solutions to the
constraint admit an expansion in terms of the perturbative parameter $\epsilon$,
the lowest order contribution in the perturbative expansion of a solution can be regarded
as a background state, in which the inhomogeneities play no dynamical role,
and which can be totally determined by its initial data
on the volume section $v=\varepsilon$. On the other hand, it can be seen that
the next contribution in the expansion of the solution satisfies the FRW constraint
equation but with a source term coming from the interaction of the inhomogeneities
and the background state. Again, we have proven that this contribution to the
solution can be determined once its initial data on the minimum
volume section are provided. Therefore, in order to construct the physical
Hilbert space, we just have to endow the space of initial data for the solutions
with a Hilbert space structure, supplying it with an inner product which, for instance, can be
selected by requiring reality conditions on a complete set of observables defined on the section
$v=\varepsilon$. The very same procedure has been carried
out in Sec.~\ref{sec:gauge_B} for the alternate gauge fixing considered in
this article.

In addition, to avoid any gauge-fixing dependence of the
possible physical outcomes of our proposals, we have introduced a family of
gauge-invariant variables. We have constructed a canonical pair of conjugate
variables out of the gauge-invariant energy and velocity perturbations (see
Ref.~\cite{bardeen} and Appendix~\ref{appendixC}). Adopting a standard Fock
description for them (namely, the so-called massless representation), we have been
able to prove that the fundamental matter perturbations in each (gauge-fixed)
reduced system and the new gauge-invariant variables are related by means of a
canonical transformation that can be implemented unitarily quantum
mechanically. Moreover, the uniqueness results of Ref.~\cite{uniq} can be immediately
applied to the gauge-invariant variables proposed in this article. Therefore,
the physics predicted by the different descriptions proposed in this work
(either based on gauge invariants or not)
is equivalent, at least as far as
standard quantum field theory in curved spacetimes is concerned.

In conclusion, we have been able to provide a full quantum description of an
inflationary universe with small inhomogeneities propagating on it, in the
context of LQC. The model is now ready to produce physical predictions, which
will be the aim of future work.

\begin{acknowledgments}
We would like to acknowledge M. Mart\'in-Benito and D. Mart\'in-de Blas,
T. Pereira, and J.M. Velhinho for
fruitful discussions. This work was supported by the research grants Nos.
MICINN/MINECO
FIS2011-30145-C03-02, MICINN FIS2008-06078-C03-03, and CPAN CSD2007-00042. J.O.
acknowledges CSIC for financial support under the grant No.\ JAE-Pre\_08\_00791,
and M.F.-M. acknowledges CSIC and the European Social Fund for support under the
grant No.\ JAEPre\_2010\_01544.

\end{acknowledgments}

\appendix
\section{Unitary transformation}\label{appendixA}

In this Appendix, we will provide a unitary transformation that cancels the
complex phases in the definition of the operator $\widehat \Omega ^2$
constructed from Eq.~\eqref{eq:quantumomega}. This operator is given by
\begin{eqnarray}
\label{eq:quantumomegasq} \nonumber
\widehat{\Omega}^2 &=&
-\hat{N}_{2\bar{\mu}}\hat{R}_+(v)\hat{R}^\dagger_-(v)\hat{N}_{2\bar{\mu}}
-\hat{N}_{-2\bar{\mu}}\hat{R}_-(v)\hat{R}^\dagger_+(v)\hat{N}_{-2\bar{\mu}}\\
&+&(|\hat{R}_+(v)|^2+|\hat{R}_-(v)|^2),
\end{eqnarray}
where
\begin{eqnarray}
R_\pm(v) &=& \frac{\pi \gamma G\hbar}{2}|v|^{1/2}|v\pm 2|^{1/2}\big[{\rm sgn}(v)+{\rm sgn}(v\pm2)\big]\nonumber\\
&& \times e^{\mp il_0\big(\frac{\bar{\mu}(v)}{2}+\frac{\bar{\mu}(v\pm2)}{2}\big)},
\end{eqnarray}
and $\bar{\mu}(v)=[\Delta/(2\pi\gamma G\hbar v)]^{1/3}$.

Given a generic function
$h(v)$, we define its complex exponentiation
\begin{eqnarray}
\hat U = e^{i l_0 \hat{h}(v)}.
\end{eqnarray}
We consider now the operator $\hat U \widehat\Omega^2 \hat
U^{-1}$ and, in particular, the $\hat{N}_{4\bar{\mu}}$ term (the contribution of
$\hat{N}_{-4\bar{\mu}}$ is given by the adjoint, and the remaining part of $\widehat\Omega^2$ is
a multiplicative operator in the $v$-representation). We can see that the complex phases cancel if, $\forall v$,
\begin{equation}
h(v+2)-h(v-2)=\frac{\bar{\mu}(v+2)}{2}+\frac{\bar{\mu}(v-2)}{2}+\bar{\mu}(v).
\end{equation}
Let us change the label $v\to v+2$ and restrict the transformation to a
superselection sector with, e.g., positive orientation of the triad, i.e.,
$v=\varepsilon+4n$ ($n\in \mathbb N$), without loss of generality. Then, we are able to
determine the explicit form of the function $h(v)$ by a recursive process:

\begin{align}\nonumber
&h(\varepsilon)=\frac{\bar{\mu}(\varepsilon)}{2},\nonumber\\
& h(\varepsilon+4)=\frac{\bar{\mu}(\varepsilon+4)}{2}+\bar{\mu}(\varepsilon+2)+\frac{\bar{\mu}(\varepsilon)}{2}+h(\varepsilon) \nonumber
\\ \nonumber
&=\frac{\bar{\mu}(\varepsilon+4)}{2}+\bar{\mu}(\varepsilon+2)+\bar{\mu}(\varepsilon),\\
\nonumber
&h(\varepsilon+8)=\frac{\bar{\mu}(\varepsilon+8)}{2}+\bar{\mu}(\varepsilon+6)
+\frac{\bar{\mu}(\varepsilon+4)}{2}+h(\varepsilon+4)\\ \nonumber
&=\frac{\bar{\mu}(\varepsilon+8)}{2}+\bar{\mu}(\varepsilon+6)+
\bar{\mu}(\varepsilon+4)+\bar{\mu}(\varepsilon+2)+\bar{\mu}(\varepsilon),\\
\label{eq:unitarymap}
&h(\varepsilon+4n)=\frac{\bar{\mu}(\varepsilon+4n)}{2}+
\sum_{j=0}^{2n-1}\bar{\mu}(\varepsilon+2j),\,\, n\in\mathbb{N}^+.
\end{align}
The constraint
$\hat{C}$, involving the operator $\widehat{\Omega}^2$, superselects sectors.
Once the analysis is restricted to a specific sector, it is natural to introduce
in it a unitary transformation of the above form. In this sector, the operator
$\widehat{\Omega}^2$ is then mapped into $\widehat{\Omega}^2_0$ [see
Eq.~\eqref{eq:quantumomega_flat}]. It is worth commenting that, for
large~$n$, the function $h(v)$ converges to the function $ v^{2/3}$ (up to a factor),
essentially because the sum in \eqref{eq:unitarymap} converges to the integral
$\int dv \,\,v^{-1/3}\sim v^{2/3}$ and the first term in that expression is
$\bar{\mu}(v)\sim o(v^{-1/3})$. Therefore, we recover the function of the
corresponding unitary map introduced in Ref.~\cite{apsv}.

\section{(Hyper-)Spherical harmonics}\label{appendixB}

We now briefly summarize the main properties of the (hyper-)spherical
harmonics $Q^n_{lm}$ on $S^3$ \cite{hh,harmon}. They form a basis (of square integrable functions
with respect to the volume element defined by the standard metric on the three-sphere) in which the
Laplace-Beltrami operator is diagonal, with a discrete and unbounded negative spectrum.
They are labeled with three integers: $n$, $l$, and $m$. The last two account
for the degeneracy of each eigenvalue of the Laplace-Beltrami operator. Their ranges are $0\leq l \leq n-1$ and $-l \leq m \leq l$.
We will display these labels explicitly only in those steps of our analysis in which
they play a relevant role.

Thus, the scalar harmonics $Q^n$ are eigenfunctions of
the Laplace-Beltrami operator on $S^3$, and specifically they satisfy
\begin{eqnarray}\label{Qn}
{{Q^n}_{|a}}^{|a}=-(n^2-1)Q^n,\quad n=1,2,3...
\end{eqnarray}
Here, the symbol $|$ denotes the covariant derivative with respect to the
round metric $\Omega_{ab}$ on the three-sphere of unit radius.
Since we deal with real scalar fields in our discussion, we choose the harmonics $Q^n$ to be {\emph{real}}.

Starting from these scalar harmonics, it is straightforward to construct a family of
vector harmonics $P^n_a$ by applying covariant derivatives:
\begin{eqnarray}
P^n_a=\frac{1}{n^2-1}{Q^n}_{|a},\quad n=2,3,4...\, .
\end{eqnarray}
These vector harmonics satisfy
\begin{eqnarray}
{{P^n_a}_{|b}}^{|b}=-(n^2-3)P^n_a, \quad {P^n_a}^{|a}=-Q^n.
\end{eqnarray}

In addition, we can also construct a family of tensor harmonics, namely, the scalar tensors
\begin{eqnarray}
Q^n_{ab}=\frac{1}{3}\Omega_{ab}Q^n,\quad  n=1,2,3...\, ,
\end{eqnarray}
and the traceless tensors
\begin{eqnarray}
P^n_{ab}=\frac{1}{n^2-1}{Q^n}_{|ab}+\frac{1}{3}\Omega_{ab}Q^n,\quad n=2,3,4...
\end{eqnarray}
These traceless tensors have the following properties
\begin{eqnarray}
{{P^n_{ab}}_{|c}}^{|c}&=&-(n^2-7)P^n_{ab}, \nonumber\\
\nonumber
{P^n_{ab}}^{|b}&=&-\frac{2}{3}(n^2-4)P^n_a, \\
{P^n_{ab}}^{|ab}&=&\frac{2}{3}(n^2-4)Q^n.
\end{eqnarray}

Finally, if we call $dv$ the integration measure on $S^3$ corresponding to the volume element
determined by the metric $\Omega_{ab}$, and we normalize the scalar harmonics so that
\begin{eqnarray}
\int dv \, Q^{n}_{lm}Q^{n'}_{l'm'}=\delta_{nn'}\delta_{ll'}\delta_{mm'},
\end{eqnarray}
it is straightforward to check that
\begin{eqnarray}
\int dv (P_a)^{n}_{lm}(P^a)^{n'}_{l'm'}&=&\frac{1}{n^2-1}\delta_{nn'}\delta_{ll'}\delta_{mm'}, \nonumber \\
\int dv (Q_{ab})^{n}_{lm}(Q^{ab})^{n'}_{l'm'}&=&\frac{1}{3}\delta_{nn'}\delta_{ll'}\delta_{mm'}, \nonumber \\
\int dv (P_{ab})^{n}_{lm}(P^{ab})^{n'}_{l'm'}&=&\frac{2}{3}\frac{n^2-4}{n^2-1}\delta_{nn'}\delta_{ll'}\delta_{mm'}.
\end{eqnarray}

\section{Bardeen potentials}\label{appendixC}

In this appendix, we provide the definitions of some relevant gauge-invariant quantities \cite{bardeen} and
discuss the dynamical relations between them. In doing
this, we use the expansion of the metric components and the matter field
given in Eqs.~\eqref{eq:lapse}-\eqref{eq:matter_field}. We
consider gauge transformations of the form $x^{\prime\mu}=x^\mu+\epsilon\xi^\mu$,
introduced in Sec.~\ref{sec:bardeen}, with the parametrization \eqref{eq:xi_param} and \eqref{eq:xi_param2}.

After a transformation of this kind, the modes of the metric can be written in terms of the original ones
and the modes of $\xi_\mu$:
\begin{eqnarray}
g_n&\mapsto&g_n+\frac{1}{e^{\alpha}} \dot{\xi}_0^n,\\ \nonumber
k_n&\mapsto&k_n-\frac{N_0}{e^{\alpha}}\left(\omega_n^2\xi_0^n+\dot\xi^n-\dot{\alpha}\xi^n\right),\\ \nonumber
a_n&\mapsto&a_n+\frac{1}{e^{\alpha}}\left(\frac{1}{3}\xi^n+\dot{\alpha}\xi_0^n\right),\\ \nonumber
b_n&\mapsto&b_n-\frac{1}{3e^{\alpha}}\xi^n,\nonumber\\ f_n&\mapsto&f_n+\frac{\dot\varphi }{e^{\alpha}}\xi_0^n .
\end{eqnarray}
Recall that the overdot stands for the derivative with respect to the conformal time $\eta$, and $\omega_n^2=n^2-1$.
With these transformation rules, the following scalar modes define
gauge-invariant quantities:

\begin{eqnarray}\label{eq:gauge_inv_phiA}
\Phi_n^A&=&g_n+\frac{1}{e^{\alpha} \omega_n^2 }\frac{d}{d\eta}\bigg(\frac{e^{\alpha}}{N_0}k_n-3e^{\alpha}\dot b_n\bigg),\\
\label{eq:gauge_inv_phiB}
\Phi_n^B&=&a_n+b_n+\frac{\dot{\alpha}}{\omega_n^2}\left(\frac{k_n}{N_0}-3\dot b_n\right), \quad\\
\label{eq:gauge_inv_Em}
\mathcal E^m_{n} &=& \frac{1}{E_0e^{2\alpha}}\left[\dot\varphi\dot f_{n}-\dot\varphi^2g_{n}+(3\dot\alpha\dot\varphi+e^{2\alpha}\tilde{m}^2\varphi)f_{n}\right] \\\label{eq:gauge_inv_vs}
v^s_{n} &=& \frac{1}{\omega_n}\bigg(\frac{\omega_n^2}{\dot\varphi}f_{n}+\frac{k_{n}}{N_0}-3\dot b_{n}\bigg),
\end{eqnarray}
where $E_0=(e^{-2\alpha}\dot \varphi^2+\tilde{m}^2\varphi^2)/2$ is proportional to the energy density of
the background scalar field. If we compare these quantities with the
gauge-invariant variables originally defined in Ref.~\cite{bardeen}, we see that
the gauge-invariant $\Phi^A$ can be identified with the potential defined in
Eq.~(3.9) of that reference. On the other hand, the Bardeen potential defined in
Eq.~(3.10) of that work corresponds to the quantity $\Phi^B$. Our invariant
energy density perturbation, $\mathcal{E}^m$, is the gauge invariant of
Bardeen's Eq.~(3.13). Finally, the gauge-invariant matter velocity of
Eq.~(3.11) in Ref.~\cite{bardeen} corresponds to $v^s$.

Let us derive now the expressions of $\mathcal{E}_n^m$ and $v^s_n$ as functions of
the canonical variables introduced in Sec.~\ref{sec:inhomog_class}. For the
gauge-invariant energy density and velocity perturbations, we need to
employ the dynamical equations in order to rewrite the momenta $\pi_{f_n}$
and $\pi_{b_n}$ in terms of the time derivative of the corresponding
conjugate variables, i.e., $\dot{f_n}$ and $\dot{b}_n$, respectively. For the
matter perturbation we obtain
\begin{equation}\label{eq:pif_n_dotf_n}
\pi_{f_n}=e^{2\alpha}\dot f_n+\pi_\varphi(3a_n-g_n).
\end{equation}
Taking this into account, a simple computation yields
\begin{equation}\label{eq:Em_matter_field}
\mathcal{E}_n^m=\frac{1}{E_0 e^{6\alpha}}\left[\pi_\varphi\big(\pi_{f_n}-3\pi_\varphi a_n\big)+
\big(e^{6\alpha}\tilde{m}^2\varphi-3\pi_\varphi\pi_\alpha\big) f_n\right].
\end{equation}
On the other hand, for the momentum conjugate to $b_n$ we have
\begin{equation}\label{eq:pib_n_dotb_n}
\pi_{b_n}=\frac{n^2-4}{n^2-1}\left(e^{2\alpha}\dot b_n-4\pi_\alpha b_n-\frac{1}{3}e^{2\alpha}\frac{k_n}{N_0}\right).
\end{equation}
Therefore, the velocity gauge-invariant perturbation can be written as
\begin{equation}\label{eq:vs_matter_field}
v^s_n=\frac{1}{\omega_n}\left[\frac{e^{2\alpha}}{\pi_\varphi}\omega_n^2 f_n-\frac{3}{e^{2\alpha}}
\left(\frac{n^2-1}{n^2-4}\pi_{b_n}+4\pi_\alpha b_n\right)\right].
\end{equation}

We now turn to the relations existing between these gauge-invariant quantities.
On the one hand, one can check that the equation of motion for the perturbation $b_n$ (see Eq.~(B11) of
Ref.~\cite{hh}) is equivalent to
\begin{eqnarray}\label{eq:gauge_inv_rel1}
\Phi^A_n+\Phi^B_n=0,\quad n=3,4\ldots
\end{eqnarray}
This equation corresponds to
Eq.~(4.4) in Ref.~\cite{bardeen}, valid for any fluid whose stress-energy tensor has a
vanishing traceless part. Another interesting relation arises from a linear
combination of the constraints $H_{|1}^n$ and $H_{\_\!\_1}^n$, given in
Eqs.~\eqref{eq:scala_const1} and \eqref{eq:diffeo_const}, respectively. One can
see that
\begin{eqnarray}\nonumber
H_{|1}^n-3\dot{\alpha}H_{\_\!\_1}^n-3H_{|0}a_n=e^{3\alpha}E_0\mathcal{E}^m_n-\frac{1}{3}e^{\alpha}(n^2-4)\Phi^B_n.
\end{eqnarray}
If the conditions $H_{|1}^n= 0$ and $H_{\_\!\_1}^n=0$ are satisfied, the last equation
reduces to Eq.~(4.3) of Ref.~\cite{bardeen}, or, equivalently,
\begin{eqnarray}\label{eq:gauge_inv_rel2}
\Phi_n^B=\frac{3e^{2\alpha}}{n^2-4}E_0\mathcal{E}^m_n.
\end{eqnarray}

Let us now define
\begin{eqnarray}
P_0&=&\frac{1}{2}\Big(\frac{\dot \varphi^2}{e^{2\alpha}}-\tilde{m}^2\varphi^2\Big),\quad
w=\frac{P_0}{E_0}, \\
c_s^2&=&\frac{dP_0}{dE_0}=1+\frac{2e^{2\alpha}\tilde{m}^2\varphi}{3 \dot{\alpha} \dot \varphi},\\
\eta_n&=&\frac{\delta P_n}{P_0}-\frac{dP_0}{dE_0}\frac{\delta E_n}{P_0}=\frac{1-c_s^2}{w}\mathcal{E}^m_n.
\end{eqnarray}
Here, $P_0$ is proportional to the pressure of the homogeneous matter field, while $\delta E_n$ and $\delta P_n$ are the modes of the energy-density and pressure perturbations. Using these formulas, one straightforwardly proves that Eq.~(4.5) of
Ref.~\cite{bardeen} can be written as
\begin{eqnarray}\label{eq:gauge_inv_rel3}
E_0\mathcal{E}^m_n=\frac{\dot \varphi^2}{e^{3\alpha}}\bigg[\frac{1}{\omega_n}\frac{d}{d\eta}(e^{\alpha}v^s_n)-e^{\alpha}\Phi^A_n
\bigg].
\end{eqnarray}

Finally, we can obtain an energy equation like Eq.~(4.8) of
Ref.~\cite{bardeen}. We start from the equation of motion of $f_n$,  which
is a second-order differential equation that can be found in Eq.~(B14) of
Ref.~\cite{hh}. If we combine it with Eqs.~\eqref{eq:gauge_inv_rel1} and
\eqref{eq:gauge_inv_rel2}, and with the equation of motion of $\alpha$, we arrive at the expression

\begin{eqnarray}\label{eq:gauge_inv_rel4}
\frac{d}{d\eta}\bigg(e^{3\alpha}E_0\mathcal{E}^m_n\bigg)=-\frac{n^2-4}{n^2-1} e^{\alpha}\dot{\varphi}^2\omega_nv^s_n.
\end{eqnarray}

\section{Gauge-invariant curvature perturbation}\label{appendixD}

Another interesting gauge-invariant quantity is the curvature perturbation,
which is defined as
\begin{eqnarray}\label{eq:curvat_inv}
\mathcal{R}_n=\Phi_n^B-\frac{\dot{\alpha}}{\omega_n}v_n^s,
\end{eqnarray}
and was originally studied by Bardeen (see the corresponding definition in
Eq.~(5.19) of Ref.~\cite{bardeen}). In the case of {\emph{spatially flat cosmologies}}, it is
common to work with a closely related quantity, the so-called Mukhanov-Sasaki
variable $v_n=z\mathcal{R}_n$, with $z=e^{\alpha}\dot{\varphi}/\dot{\alpha}$. In such situations,
the power spectrum of primordial perturbations can be easily derived using the
Mukhanov-Sasaki variable.

In our case, this gauge invariant has different expressions in terms of the
canonical variables employed in the two distinct gauge fixings considered here.

\subsubsection{Longitudinal gauge}

In order to compute $\mathcal{R}_n$ in terms of our fundamental fields $\bar f_n$ and
$\bar\pi_{f_n}$, we recall first that the potential $\Phi_n^B$
is related with $\mathcal{E}^m_n$ by means of Eq.~\eqref{eq:gauge_inv_rel2}.
Applying the results of Sec.~\ref{sec:bardeen}, it is then easy to obtain that
\begin{equation}\label{eq:curvat_inv_gaugeA}
\mathcal{R}_n=\frac{3\pi_\varphi}{e^{3\alpha}(n^2-4)}(\bar\pi_{f_n}+\chi_A\bar f_n)
+\frac{\pi_\alpha}{e^{\alpha}\pi_\varphi}\bar f_n.
\end{equation}
The variable $\chi_A$ was defined in Eq.~\eqref{eq:chiA}. Clearly, in the
large $n$ limit, the main contribution to the previous expression comes from the last factor, proportional
to $\bar f_n$.

\subsubsection{Gauge fixing $a_n=b_n=0$}

This gauge provides a simple relation
between the curvature perturbation and the modes $\bar f_n$. Actually, using the
definitions~\eqref{eq:curvat_inv}, \eqref{eq:gauge_inv_phiB}, and
\eqref{eq:gauge_inv_vs}, a straightforward computation yields
\begin{eqnarray}\label{eq:curvat_inv_gaugeB}
\mathcal{R}_n=\frac{\pi_\alpha}{e^{\alpha}\pi_\varphi}\bar f_n.
\end{eqnarray}
Therefore, in this gauge, where the spatial geometry is
homogeneous, the unique contribution to the curvature perturbation comes
essentially from the perturbation of the scalar field.

Moreover, in the ultraviolet limit, the
form of the curvature perturbation coincides formally with that of
the longitudinal gauge, given by Eq.~\eqref{eq:curvat_inv_gaugeA}.
As a consequence, for the two studied gauges, the
Mukhanov-Sasaki variable $v_n$ either coincides with $\bar f_n$ (gauge $a_n=b_n=0$)
or converges to it (longitudinal gauge).


\begin{thebibliography}{M}
\bibitem{note} For instance, an inner product can be selected by
choosing a complete set of real observables and requiring that their complex conjugation relations
be promoted to adjointness relations between the operators that represent them
(see Ref. \cite{reacon}).
\bibitem{comment1b} Although we use the same notation for the integers labeling the
homogeneous volume sections $v=\varepsilon+4n$ and the eigenvalues of the
Laplace-Beltrami operator $\omega_n^2=n^2-1$, this should lead to no confusion since the
actual meaning of the label must be clear from the context in which it is employed in the manuscript.
\bibitem{comment2} As we will discuss in Sec.~\ref{sec:bardeen}, even when the
results of Ref.~\cite{uniq} pick up a particular pair of canonical
fields among those related by the considered class of time-dependent canonical transformations,
more general canonical transformations can in principle be implemented
unitarily as well, e.g., in cases in which we may allow them to have a suitable mode dependence.
\bibitem{comment3} After the application of the unitary transformation $\hat U$
defined in Sec.~\ref{sec:hom_quant}, the operator $\widehat{\Omega}^{2}$
transforms into $\widehat{\Omega}_0^{2}$ [see Eq.~\eqref{eq:flat_omega}], which is
an essentially self-adjoint operator, with a continuous, positive, and
non-degenerate spectrum. See the discussion in Sec.~\ref{sec:hom_quant} and
Refs.~\cite{apsv,mmo} for additional details.
\end{thebibliography}
\end{document}